\documentclass{eptcs}
 \nonstopmode

 \newcommand{\Comment}[1]{}

 \newif\ifmycolour \mycolourfalse

 \usepackage{pstricks} 
 \usepackage{qsymbols}
 \usepackage{mathpazo}
 \usepackage{times}
 \usepackage{breakurl}
 \usepackage{derivation}

 \newenvironment{proof}{ \vspace*{1mm} \item[\hskip \labelsep {\it Proof.}] \ignorespaces \rm}{}

 \input{Macros}
\newtheorem{definition}{Definition}[section]{\bf}{\rm}
 {\it}{\rm}
 {\it}{\rm}
 {\it}{\it}
 {\it}{\it}
 \newtheorem{corollary}[definition]{Corollary}{\it}{\it}
 {\it}{\rm}
 \newtheorem{lemma}[definition]{Lemma}{\it}{\it}
 {\it}{\it}
 \newtheorem{proposition}[definition]{Proposition}{\it}{\it}
 \newtheorem{theorem}[definition]{Theorem}{\bf}{\it}

\begin{document}

 \title{Characterisation of Strongly Normalising $`l`m$-Terms}
 \author{%
Steffen van Bakel
 \institute{Imperial College London \\ London, UK}
 \email{svb@doc.ic.ac.uk}
 \and
Franco Barbanera
 \institute{Universit\`a di Catania \\ Catania, Italy}
 \email{barba@dmi.unict.it}
 \and
Ugo de'Liguoro
 \institute{Universit\`a di Torino \\ Torino, Italy}
 \email{deliguoro@di.unito.it}
}

 \def\titlerunning{Characterisation of Strongly Normalising $`l`m$-Terms}
 \def\authorrunning{van Bakel, Barbanera, and de'Liguoro}

 \date{}

 \bibliographystyle{eptcs}

 \maketitle

 \begin{abstract} 
We provide a characterisation of strongly normalising terms of the $`l`m$-calculus by means of a type system with intersection and product types.
The presence of the latter and a restricted use of the type $`w$ enable us to represent the particular notion of continuation used in the literature for the definition of semantics for the $`l`m$-calculus. 
This makes it possible to lift the well-known characterisation property for strongly-normalising $`l$-terms - that uses intersection types - to the $`l`m$-calculus. 
From this result an alternative proof of strong normalisation for terms typeable in Parigot's propositional logical system follows, by means of an interpretation of that system into ours. 
 \end{abstract}

 \section*{Introduction}

Parigot's $`l`m$-calculus \cite{Parigot'92} is an extension of the $`l$-calculus \cite{Church'36,Barendregt'84} that was first introduced in \cite{Parigot'92} to express a notion of (confluent) computation with classical proofs in Gentzen's \emph{sequent calculus} {\LK}.
That calculus was introduced in \cite{Gentzen'35} as a logical system in which the rules only introduce connectives (but on either side of a sequent), in contrast to \emph{natural deduction} (also introduced in \cite{Gentzen'35}) which uses rules that introduce or eliminate connectives in the logical formulae.
Natural deduction normally derives statements with a single conclusion, whereas {\LK} allows for multiple conclusions, deriving sequents of the form $ { A_1,\ldots,A_n } \vdash { B_1,\ldots,B_m }$, where $A_1,\ldots,A_n$ is to be understood as $A_1 {\wedge}\ldots{\wedge} A_n$ and $B_1,$ $\ldots,B_m$ is to be understood as $B_1 {\vee}\ldots{\vee} B_m$.

With $`l`m$, Parigot created a multi-conclusion typing system that is, in fact, based on a mixture of Gentzen's two approaches: the system is a natural deduction system that has \emph{introduction} and \emph{elimination} rules, but derivable statements have the shape $\derLmu `G |- M : A | `D $, where $A$ is the main conclusion of the statement, expressed as the \emph{active} conclusion.
Here $`D$ contains the alternative conclusions, consisting of pairs of Greek characters and types; the left-hand context $`G$, as usual, contains pairs of Roman characters and types, and represents the types of the free term variables of $M$.
This yields a logic with \emph{focus} where the main conclusion is the focus of the proof; derivable judgements correspond to provable statements in \emph{minimal classical logic} \cite{Ariola-Herbelin'03}.
In addition to the normal $`l$-calculus reduction rules, Parigot needed to express that the focus of the derivation (proof) changes; he therefore added \emph{structural} rules, where elimination takes place for a type constructor that appears in one of the alternative conclusions (the Greek variable is the name given to a subterm).
This is achieved by extending the syntax with two new constructs $[`a]M$ and $`m`a.M$ that act as witness to \emph{deactivation} and \emph{activation}, which together move the focus of the derivation.
The collection of reduction rules Parigot defined are carefully engineered to yield a confluent reduction system; normally, systems based on classical logic are not confluent, as is the case for example for the Symmetric $`l$-calculus \cite{Barbanera-Berardi'96}, $\lmmt$ \cite{Curien-Herbelin'00}, and $\X$ \cite{Bakel-Lescanne-MSCS'08}. 

In spite of being motivated by classical logic, the $`l`m$-calculus itself is type free. 
As a consequence there exist more terms than proofs, and properties of pure $`l`m$-terms have been extensively studied (see e.g.~\cite{Saurin'08,Herbelin-Saurin'10,Saurin'10}).
In particular, among them there are perfectly meaningful terms that do not correspond to any proof, like fixed-point constructors for example. 
The basic idea here to turn non-constructive proofs into algorithms is to add a form of continuation by means of \emph{names} and $`m$-\emph{abstraction} to capture (a notion of) control. 
However, continuations introduce a great deal of complexity to the calculus' semantics and inspired by the results proven in \cite{Bakel-ITRS'10} we decided to explore the possibility of defining \emph{filter} semantics for $`l`m$. 
Starting from Streicher and Reus' denotational semantics of $`l`m$ in \cite{Streicher-Reus'98}, in \cite{Bakel-Barbanera-Liguoro-TLCA'11} we have introduced an intersection type assignment system that induces a filter model. 
This, essentially, is a logical description of the domain-theoretic model of \cite{Streicher-Reus'98}, with the advantage of providing a formal tool to reason about the meaning of terms. 

One of the main results for $`l`m$, proved in \cite{Parigot'97}, states that all $`l`m$-terms that correspond to proofs of second-order natural deduction are strongly normalising; the reverse of this property does not hold for Parigot's system, since there, for example, not all terms in normal form are typeable.

The full characterisation of strong normalisation ($M$ is strong normalising if and only if $M$ is typeable) is a property that is shown for various intersection systems for the $`l$-calculus, and towards the end of \cite{Bakel-Barbanera-Liguoro-TLCA'11} we conjectured that in an appropriate subsystem we would be able to type exactly all strongly normalising $`l`m$-terms as well.
The first to state the characterisation result was Pottinger \cite{Pottinger'80} for a notion of type assignment similar to the intersection system of \cite{Coppo-Dezani'78,Coppo-Dezani-Venneri'80}, but extended in that it is also closed for $`h$-reduction, 
and is defined without the type constant $`w$.
However, to show that all typeable terms are strongly normalisable, \cite{Pottinger'80} only \emph{suggests} a proof using Tait's computability technique \cite{Tait'67}.
A detailed proof, using computability, in the context of the $`w$-free BCD-system \cite{BCD'83} is given in \cite{Bakel-TCS'92}; to establish the same result saturated sets are used by Krivine in \cite{Krivine-book'93} (chapter 4), in Ghilezan's survey \cite{Ghilezan'96}, and in \cite{Bakel-ACM'11}.

The converse of that result, the property that all strongly normalisable terms are typeable has proven to be more elusive: it has been claimed in many papers but not shown in full (we mention \cite{Pottinger'80,Bakel-TCS'92,Ghilezan'96}); in particular, the proof for the property that type assignment is closed for subject expansion (the converse of subject reduction) is dubious.
Subject expansion can only reliably be shown for \emph{left-most outermost} reduction, which is used for the proofs in \cite{Krivine-book'93,Bakel-NDJFL'04,Bakel-ACM'11}, and our result follows that approach as well.

In the full system of \cite{Bakel-Barbanera-Liguoro-TLCA'11}, all terms are typeable with $`w$ and this clearly interferes with the termination property.
However, the problem we face is slightly more complex than straightforwardly removing $`w$, as done in \cite{Bakel-TCS'92,Bakel-NDJFL'04}.
In the model (for details, see \cite{Bakel-Barbanera-Liguoro-TLCA'11}) a \emph{continuation} is an infinite tuple of terms, which is typed in the system by (a finite intersection of) types $`k = `d_1\prod\cdots`d_k\prod`w$ for some $k>0$, where the leading $`d_1,\ldots,`d_k$ encode the information about the first $k$ terms in the tuple, while the ending $`w$ represents the lack of information about the remaining infinite part.
This implies that, for our system for $`l`m$, we cannot remove $`w$ completely. 
To solve this problem, 
we first restrict types to those having $`w$ only as the final part of a product type; we then suitably modify the standard interpretation of intersection types, adapting Tait's argument in such a way that the semantics of $`k$ is the \emph{set of all finite tuples} $\Vec{L}$ (called {\em stacks}) of strongly normalising terms that begin with $k$ terms $L_1\ldots,L_k$ that belong to the interpretations of, respectively, $`d_1,\ldots,`d_k$. 
For this restricted system, we will show that typeability characterises strong normalisability for $`l`m$-terms.

As a consequence of our characterisation result we also obtain an alternative proof of Parigot's termination result \cite{Parigot'97} (for the propositional fragment), by interpreting ordinary types into our intersection types and proving that the translation preserves derivability from Parigot's system to ours.

 \paragraph{Outline of this paper.}
In Section \ref{sec:lambdaMu}, we will briefly recall Parigot's untyped $`l`m$-calculus \cite{Parigot'92}. 
After defining appropriate sets of types in \ref{sec:system}, a pre-order over types, and our typeing system in Section~\ref{subsec:typImpSN}, we will show that typeability implies strong normalisation. 
The opposite implication, proved in Section~\ref{sub:NormType}, will complete our main results. 
The alternative proof of Parigot's theorem for the propositional fragment will be developed in Section \ref{sec:Parigot}, and we finish by giving concluding remarks.

 \section{The $`l`m$-calculus}\label{sec:lambdaMu}

In this section we present Parigot's pure $`l`m$-calculus as introduced in \cite{Parigot'92}, slightly changing the notation.

 \begin{definition} [Term Syntax \cite{Parigot'92}] \label{def:terms}

 \begin{enumerate}
 \item
The sets $\Terms$ of {\em terms} and $\Commands$ of {\em commands} are defined inductively by the following grammar (where $x \ele \TVar$, a set of \emph{term variables}, and $`a \ele \CVar$, a set of \emph{names}, both denumerable):
 \[ \begin{array}{rll@{\hspace{1cm}}l}
M,N & ::= &x \mid `l x.M \mid MN \mid `m`a.C & \textrm{(terms)}
 \\ [1mm] 
C & ::= & [`a] M & \textrm{(commands)} 
 \end{array} \]

 \item
We call $ \Vec{L} \equiv L_1\cons \dots \cons L_k$ a \emph{stack of terms}; we denote the set of all finite (possibly empty) stacks of terms by $ \Terms^*$, and write $\epsilon$ for the empty stack.
If $M \ele \Terms$ and $ \Vec{L} \equiv L_1\cons \dots \cons L_k$ then $M\cons \Vec{L} \equiv M\cons L_1\cons \dots \cons L_k \ele \Terms^*$, while we define $M (P\cons \Vec{L} ) \ByDef MP\Vec{L} $, so $M \Vec{L} \equiv M L_1 \cdots L_k$.

 \end{enumerate}
 \end{definition}
We will often speak of a stack rather than a stack of terms.
For convenience of notation, for $ \Vec{L}= L_1 \cons \dots \cons L_k \ele \Terms^*$, we introduce the notation:
 \[ \begin{array}{rcl}
M \strSubvec[ `a <= L ] &\ByDef& M \strSub[ `a <= L_1 ] \strSub[ `a <= L_2 ] \dots \strSub[ `a <= L_n ]
 \end{array} \]
when each $L_i$ does not contain $`a$.
In particular, $M \strSub[ `a <= {\epsilon} ] \ByDef M$.
Notice that, by definition of structural substitution,
 \[ \begin{array}{lclclcl}
[`a]M \strSub[ `a <= \Vec{L} ] 
 & \ByDef &
[`a]M \strSub[ `a <= L_1 ] \strSub[ `a <= L_2 ] \dots \strSub[ `a <= L_n ] 
 & \ByDef &
[`a](M\strSub[ `a <= {\Vec{L}} ]) \Vec{L} 
 \end{array} \]

As usual, we consider $`l$ and $`m$ to be binders; we adopt Barendregt's convention on terms, and will assume that free and bound variables are different; the sets $\fv(M)$ and $\fn(M)$ of, respectively, \emph{free variables} and \emph{free names} in a term $M$ are defined in the usual way. 

 \begin{definition} [Substitution \cite{Parigot'92}] \label{def:substitution}
Substitution takes two forms:
 \[ \begin{array}{l@{\hspace{2mm}}cll}
 \textit{term substitution:} & M[N/x] & \textrm{($N$ is substituted for $x$ in $M$, avoiding capture)} 
 \\
 \Comment{
\textit{renaming:} & C[`a/`b ] & \textrm{(every free occurrence of $`b $ in $C$ is replaced by $`a$)} \\
}
 \textit{structural substitution:} & T\StrSub{`a}{L} & 
			\textrm{(every subterm $[`a]N$ of $M$ is replaced by $[`a]NL$)}
 \end{array} \]
where $M,N,L \ele \Terms$, $C \ele \Commands$ and $T \ele \Terms \cup \Commands$.
More precisely, $T\StrSub{`a}{L}$ is defined by:
 \[ \begin{array}{rcll}
([`a]M)\StrSub{`a}{L} &\ByDef& [`a](M\StrSub{`a}{L})L 
	\\
([`b ]M)\StrSub{`a}{L} & \ByDef & [`b ]M\StrSub{`a}{L} \dquad \textrm{if $`a\neq`b $}
	\\
(`m`b .C)\StrSub{`a}{L} &\ByDef & `m`b .C\StrSub{`a}{L} 
\\ 
x\StrSub{`a}{L} & \ByDef & x 
	\\
(`l x.M)\StrSub{`a}{L} & \ByDef & `l x.M\StrSub{`a}{L} 
	\\
(MN)\StrSub{`a}{L} & \ByDef & (M\StrSub{`a}{L})(N\StrSub{`a}{L}) 
 \end{array} \]

 \end{definition}

 \begin{definition}[Reduction \cite{Parigot'92}]\label{def:reduction}
The reduction relation $M \reduces N$, where $M,N \ele \Terms$, is defined as the compatible closure of the following rules:
 \[ \begin{array}{r@{\quad}rll@{\dquad}l}
(`b ): & (`l x.M)N & \reduces & M[N/x] & (\textit{logical reduction}) \\
(`m): & (`m`b .C)N & \reduces & `m`b .C\StrSub{`b }{N} & (\textit{structural reduction}) \\
 \end{array} \]
 \end{definition}

 \section{Characterisation of Strong Normalisation} \label{sec:character}

In this section we will show that we can characterise strong normalisation for pure $`l`m$-terms completely through a notion of intersection typeing which employs product types and a restricted use of the type $`w$.


 \subsection{The type system}\label{sec:system}

As mentioned in the introduction, our characterisation can be carried out by means of a precisely tailored version of the type system we presented in \cite{Bakel-Barbanera-Liguoro-TLCA'11}.
The types of our system will be formed by means of the $\arrow$, $\prod$, and $\inter$ type constructors over a single base type $`n$.%
\footnote{In \cite{Bakel-Barbanera-Liguoro-TLCA'11}, more base types are used, but for our present purposes one suffices.}

 \begin{definition}[Types]\label{def:restrTypes}
The sets $\Lang_D$ of {\em term types} and $\Lang_C$ of {\em continuation-stack types} are
defined inductively by the following grammar, where $`n$ is a type constant:
 \[ \begin{array}{l@{\quad}rcl@{\quad}l}
 \Lang_D : & `d &::=& `n \mid `w\arrow`n \mid `k\arrow`n \mid `d \inter `d 
	& 
	\\ [1mm]
 \Lang_C : & `k &::= & `d\prod `w \mid`d\prod `k \mid `k \inter `k
	& 
 \end{array} \]
(we will call the types $`d\prod `w$ and $`d\prod `k $ also \emph{product types}).
We define the set $\Lang$ of {\em types} as $\Lang = \Lang_D \cup\Lang_C $ and let $`s$, $`t$, $`r$, etc.~range over $ \Lang$.
 \end{definition}

Notice that an important feature of our system is the absence of $`w$ as a proper type (and, consequently, the absence of its corresponding typeing rule); notice that we have not removed $`w$ completely, since it always occurs at the very end of any product type in order to represent the (unspecified) last part of a continuation stack.

 \begin{definition}\label{def:leq-minus} \label{leq-times-ext}
The relations $\leqD $ and $\leqC $ are the least pre-orders over $\Lang_D$ and $\Lang_C$,
respectively, such that:
 \[ \begin{array}{c}
 \Inf	{\sigma \inter `t \leqA \sigma }
 \dquad
 \Inf	{\sigma \inter `t \leqA `t }
 \dquad
 \Inf	{`n \leqD `w\arrow`n }
 \dquad
 \Inf	{ `w\arrow`n \leqD `n }
 \dquad
 \Inf	{`d _1\prod `d _2\prod `w \leqC `d _1 \prod `w } 
 \\ [3mm]
 \Inf	{(`d _1\prod `w )\inter(`d _2\prod `k ) \leqC (`d _1\inter`d _2)\prod `k }
 \dquad
 \Inf	[ `k _1, `k _2\not\equiv`w ]
	{(`d _1\prod `k _1) \inter (`d _2\prod `k _2) \seqC (`d _1\inter`d _2)\prod ( `k _1\inter `k _2) }
 \\ [4mm]
 \Inf	{`d _1\leqD `d _2 }{`d _1\prod `w \leqC `d _2\prod `w }
 \dquad
 \Inf	{`d_1\seqD `d _2 \quad `k _1\seqC `k _2 }{`d _1\prod `k _1 \seqC `d _2\prod `k _2 }
 \dquad
 \Inf	{\sigma \leqA `t _1 \quad \sigma \leqA `t _2}{\sigma \leqA `t _1\inter `t _2}
 \dquad
 \Inf	{`k _2 \seqC `k _1 }{`k _1\arrow`n \seqD `k _2\arrow`n }
 \end{array} 
 \]
where $A$ is either $D$ or $C$.
As usual, we define ${=_A} \ByDef {\leqA\cap\geq_A}$. 
 \end{definition}
For convenience of notation, in the following the subscripts $D$ and $C$ on $\leq$ are normally omitted. \\

 \noindent
The pre-orders in Definition~\ref{def:leq-minus} are a restriction to $\Lang$ of the pre-orders defined in \cite{Bakel-Barbanera-Liguoro-TLCA'11}.
We point out that, in the system defined in that paper, the inequality $`d _1\prod `d _2\prod `w \leq `d _1\prod `w $ is derivable. 
In fact, in \cite{Bakel-Barbanera-Liguoro-TLCA'11} we had $`w =_C `w\prod `w$ and hence
$`d _1\prod `d _2\prod `w \leq `d _1\prod `w\prod `w = `d _1\prod `w $. In the present system, instead, $`w\not\in\Lang_D $ so that $`d _1\prod `w\prod `w\not\in\Lang_C $, and therefore this inequality has to be explicitly postulated above.


The notions of \emph{basis} (\emph{variable context}), denoted by $`G$, $`G'$, \ldots, and \emph{name context}, denoted by $`D$, $`D'$, \ldots, are defined in the standard way as, respectively, mappings of a finite set of term variables to types in $\Lang_D$, and of a finite set of names to types in $\Lang_C$, represented for convenience as sets of statements on variables and names (we call these assumptions).
Below we shall write $`G, x{:}`d$ for $`G \cup \Set{x{:}`d}$ where $x\not\in\Dom(`G)$; similarly for $`a{:} `k ,`D$ (note that the order in which variable and name assumptions are listed in the rules is immaterial).

 \begin{definition}[Typeing System]
 \begin{enumerate}
 \item
A \emph{judgement} in our system has the form $ \derLmu `G |- M : `d | `D $, where $`G$ is a basis, $M\in\Terms$, $`d\in\Lang_D$ and $`D$ is a name context.

 \item
We define \emph{typeing} for pure $`l`m$-terms (in $\Terms$) through the following natural deduction system:
 \[ \begin{array}{rl@{\quad}rl}
(\Ax) : &
 \Inf	{\derLmu `G, x{:}`d |- x{:}`d | `D }
&
(`m) : &
 \Inf	{\derLmu `G |- M : `k'\arrow`n | `a{:}`k,
 `D
	}{\derLmu `G |- `m`a.[`b]M : `k\arrow`n | `b{:}`k ',`D }
 \quad
 \Inf	{\derLmu `G |- M : `k\arrow`n | `a{:}`k,`D
	}{\derLmu `G |- `m`a. [`a]M : `k\arrow`n | `D }
 \\ [6mm]
(\Abs): &
 \Inf	{\derLmu `G, x{:}`d |- M : `k\arrow`n | `D
	}{\derLmu `G |- `lx.M : `d\prod `k\arrow`n | `D }
&
(\App) : &
 \Inf	{\derLmu `G |- M : `d\prod`k\arrow`n | `D
	 \quad
	 \derLmu `G |- N : `d | `D
	}{\derLmu `G |- MN : `k\arrow`n | `D }
 \\ [6mm]
(\leq) : &
 \Inf	[`d \leq `d']
	{\derLmu `G |- M : `d | `D \quad
	}{\derLmu `G |- M : `d' | `D }
&
(\inters) : &
 \Inf	{\derLmu `G |- M : `d | `D \quad \derLmu `G |- M : `d' | `D
	}{\derLmu `G |- M : `d\inter`d' | `D }
 \end{array} \]
where $ `k $ in rules $(\Abs)$ and $(\App)$%
\footnote{We use $(\App)$ and $(\Abs)$ to name the rules concerning $`l$-abstraction and application, rather than the more usual $(\arrowI)$ and $(\arrowE)$, since in our system there is no introduction or elimination of the $\arrow$ type constructor.}
is either a type in $\Lang_C $ or $`w$.

 \item
We write $ \derLmu `G |- M : `d | `D $ whenever there exists a derivation built using the above rules that has this judgement in the bottom line, and $\Der :: \derLmu `G |- M : `d | `D $ when we want to name that derivation.
We write $ \derLmu {} |- M : `d | `D $ when the variable context is empty, and $ \derLmu `G |- M : `d | {} $ when the name context is.

 \end{enumerate}
 \end{definition}

Note that we use a single name, $(`m)$, for the two rules concerning $`m$-abstraction; which is the one actually used will always be clear from the context. 

We extend Barendregt's convention to judgements $ \derLmu `G |- M : `d | `D $ by seeing the variables that occur in $`G$ and names in $`D$ as binding occurrences over $M$ as well; in particular, we can assume that no variable in $`G$ and no name in $`D$ is bound in $M$.

 \begin{definition}
 \begin{enumerate}

 \item
The relation $\leq$ is naturally extended to bases as follows:
 \[ \begin{array}{rcl}
`G' \leq `G & \textit{iff} & x{:}`d\ele `G \Then \Exists x{:}`d'\ele `G' \Pred [ `d' \leq`d ].
 \end{array} \]
The $\leq$ relation on name contexts is defined in the same way.

 \item
Given two bases $`G_1$ and $`G_2$, we define the basis $`G_1\inters `G_2$ as follows:
 \[ \begin{array}{rcrl}
`G_1\inters`G_2 &\ByDef& \Set{x{:}`d_1\inters `d_2 \mid x{:}`d_1 \ele `G_1 \And x{:}`d_2 \ele `G_2 } & \Union \\
&& \Set {x{:}`d \mid x{:}`d \ele `G_1 \And x \notele \Dom(`G_2) } & \Union \\
&& \Set {x{:}`d \mid x{:}`d \ele `G_2 \And x \notele \Dom(`G_1) } 
 \end{array} \]

 \item
The name context $`D_1\inters`D_2$ is constructed out of $`D_1$ and $`D_2$ in a similar way.
 \end{enumerate}

 \end{definition}

Trivially, $\Dom(`G_1\inters`G_2) = \Dom(`G_1) \union \Dom(`G_2)$ and $\Dom(`D_1\inters`D_2) = \Dom(`D_1) \union \Dom(`D_2)$.
Moreover, it is straightforward to show that:

 \begin{proposition} \label{prop:interleq} 
$`G_1\inters`G_2 \leq `G_{i}$ and $`D_1\inters`D_2 \leq `D_{i}$ for $i=1,2$.
 \end{proposition}

\Comment{
 \subsection{Properties of the type assignment system}\label{sub:systemProp}
}

We can also show that \emph{Weakening} and \emph{Strengthening} rules are implied by the system:

 \begin{lemma} [Weakening and Strengthening] \label{lem:weakening} \label{restricted weakening} \label{restricted strengthening}
The following rules are admissible%
\footnote{We should perhaps point out that Barendregt's convention, extended to judgements as we do here, is essential for the correctness of this result.
By writing $\derLmu `G' |- M : `d | `D' $, we assume that $`G'$ and $`D'$ do not contain statements for variables and names that occur bound in $M$, so we do not allow contexts to be weakened by statements concerning bound names or variables.
As a counter example, take $ \derLmu {} |- `m`a.[`a]`lx.x : (`k\arrow`n)\arrow`k\arrow`n | {} $ and $`G_2 = x{:}`d$, $`D_2 = `a{:}`k$; we cannot derive $ \derLmu x{:}`d |- `m`a.[`a]`lx.x : (`k\arrow`n)\arrow`k\arrow`n | `a{:}`k $. 

This is also the case for systems for the $`l$-calculus; in past papers it has been claimed that, if $ \derL `G_1 |- M : A $ and $ \derL `G_2 |- N : B $ (without any restrictions), then also $ \derLmu `G_1\inter`G_2 |- M : A $ and $ \derLmu `G_1\inter`G_2 |- N : B $.
This is incorrect for the same reason: take 
$\derL {} |- `ly.y : A\arrow A $ and $ \derLmu y{:}(A\arrow A)\inter A\arrow A |- yy : A $; we \emph{cannot} derive $ \derL y{:}(A\arrow A)\inter A\arrow A |- `ly.y : A\arrow A $.
}:
 \[ \begin{array}{rl}
(\Weak): &
 \Inf	[ `G' \leq `G,`D' \leq `D ]
	{\derLmu `G |- M : `d | `D
	}{\derLmu `G' |- M : `d | `D' }
 \\ [5mm]
(\textit{S}): &
 \Inf	
	[`G' = \Set {x{:}`d\ele`G \mid x \ele \fv(M) }, ~ `D' = \Set {`a{:}`k\ele`D \mid `a \ele \fn(M) }]
	{\derLmu `G |- M : `d | `D
	}{\derLmu `G' |- M : `d | `D' }
 \end{array} \]

 \end{lemma}

 \Comment{
 \begin{lemma}[Strengthening Lemma]\label{lem:strength} \hfill
 \begin{enumerate}
 \item \label{lem:strength i}
	$ \derLmu `G,x{:}`d_1 |- M : `d | `D \And `d_2\leq`d_1 \Then
	\derLmu `G,x{:}`d_2 |- M : `d | `D $.
 \item \label{lem:strength ii}
	$ \derLmu `G |- M : `d | `a{:}`k_1,`D \And `k_2\leq `k_1 \Then
	\derLmu `G |- M : `d | `a{:}`k_2,`D $.xxx
 \end{enumerate}

 \begin{proof}
 \begin{description}
 \item [\ref{lem:strength i}] By induction on the structure of derivations 
In case $(\mbox{\sf \it Var})$ we have $M\equiv x$ and $`d \equiv `d_1$, so we just replace the axiom by
 \[ \begin{array}{rl}
(\leq): &
 \Inf	[`d_2\leq`d_1]
	{\derLmu `G,x{:}`d_2 |- x{:}`d_2 | `D \quad
	}{\derLmu`G, x{:}`d_2 |- x{:}`d_1 | `D }
 \end{array} \]
All the other cases are immediate by induction.

 \item [\ref{lem:strength ii}]
By induction on the structure of derivations. 
The only non trivial case is $(`m_2)$, when the derivation ends by:
 \[
 \Inf	[`m_2]
	{\derLmu `G |- M' : `k_1\arrow`n | `a{:}`k_1,`b{:}`k,`D
	}{\derLmu `G |- `m`b. [`a]M' : `k\arrow`n | `a{:}`k_1,`D }
 \]
where $M \equiv `m`b. [`a]M'$ and $`d \equiv `k\arrow`n$.
Since $ `k_2\leq `k_1$ implies $ `k_1\arrow`n \leq `k_2\arrow`n$, we replace the last inference by:
 \[
 \Inf	[`m_2]
	{\Inf	[`k_1\arrow`n \leq `k_2\arrow`n]
		{\derLmu `G |- M' : `k_1\arrow`n | `a{:}`k_2,`b{:}`k ,`D
		}{\derLmu `G |- M' : `k_2\arrow`n | `a{:}`k_2,`b{:}`k ,`D }
	}{\derLmu `G |- `m`b. [`a]M' : `k\arrow`n | `a{:}`k_2,`D }
 \]
where $`G |- M' : `k_1\arrow`n | `a{:}`k_2, `b{:}`k ,`D $ is derivable by induction.\QED

 \end{description}
 \end{proof}
 \end{lemma}
}

The above lemma and Proposition \ref{prop:interleq} lead immediately to the following:

 \begin{corollary} \label{lem:contextInters}
If $ \derLmu `G_1 |- M : `d | `D_1 $ then for any $`G_2$, $`D_2$:
$ \derLmu `G_1\inters `G_2 |- M : `d | `D_1\inters`D_2 $.
 \end{corollary}
Notice that, by Barendregt's convention, the variables in $`G_2$ and names in $`D_2$ are not bound in $M$.

The following substitution results can be proved along the lines of similar ones in \cite{Bakel-Barbanera-Liguoro-TLCA'11}:

 \begin{lemma}[Substitution Lemma] \label{lem:substitution} 
 \begin{enumerate}
 \item \label{lem:substitution i}
$ \derLmu `G |- M[N/x] : `d | `D $ with $x\ele\fv(M)$, if and only if there exists $`d' $ such that $ \derLmu `G |- N : `d' | `D $ and $ \derLmu `G, x{:}`d' |- M : `d | `D $.
	
 \item \label{lem:substitution ii}
$ \derLmu `G |- M \strSub[ `a <= L ] : `d | `a{:}`k,`D $ with $`a\ele\fn(M)$, if and only if there exists $`d'$ such that $ \derLmu `G |- L : `d' | `D $, and $ \derLmu `G |- M : `d | `a{:}`d'\prod `k,`D $. \end{enumerate}
 \end{lemma}

 \Comment
{

The following lemma is convenient for the proof of Lemma~\ref{lem:redexExpBeta}.

 \begin{lemma}\label{lem:EEadmissible}
The following rule is admissible:
 \[
(@): \quad
 \Inf	{\derLmu `G,x{:}`d |- M : `d' | `D \quad \derLmu `G |- N : `d | `D
	}{\derLmu `G |- (`lx.M)N : `d' | `D }
 \]

 \begin{proof}
By Barendregt's convention we can assume that $x\notele \fv(N)$ and $x\not\in\Dom(`G)$.
Now $`d' = \inters_n `k_i\arrow`n$ for some $n$ and, using $(\leq)$, from $ \derLmu `G,x{:}`d |- M : \inters_n`k_i\arrow`n | `D $ we can derive $ \derLmu `G,x{:}`d |- M : `k_i\arrow`n | `D $ for all $ i \leq n $.
Now we can construct:
 \[
 \Inf	[\App]
	{\Inf	[\Abs]
		{\derLmu `G,x{:}`d |- M : `k_i\arrow`n | `D
		}{\derLmu `G |- `lx.M : `d\prod `k_i\arrow`n | `D }
	 \quad
	 \derLmu `G |- N : `d | `D
	}{\derLmu `G |- (`lx.M)N : `k_i\arrow`n | `D }
 \]
for all $i \leq n$; we obtain $ \derLmu `G |- (`lx.M)N : `d' | `D $ by repeated application of rule $(\inters)$.
 \end{proof}
 \end{lemma}
}

 \subsection{Typeability implies Strong Normalisation}\label{subsec:typImpSN}
In this subsection we will show that -- as can be expected of a well-defined notion of type assignment that does not type recursion and has no general rule that types all terms -- all typeable terms are strongly normalising.
Such a property does not hold for the system in \cite{Bakel-Barbanera-Liguoro-TLCA'11} where, in fact, by means of types not allowed in the present system, it is possible to type the fixed-point constructor $`lf.(`lx.f(xx))(`lx.f(xx))$ in a non-trivial way, as shown by the following derivation:

 \[
 \Inf	[\Abs]
	{\Inf	[\App]
		{\Inf	[\Abs]
			{\Inf	[\App]
				{\Inf	[\Ax]
					{\derLmu f{:}`w\prod `w\arrow `n,x{:}`w |- f : `w\prod `w\arrow `n | {} }
				 \quad
				 \Inf	[`w]
					{\derLmu f{:}`w\prod `w\arrow `n,x{:}`w |- xx : `w | {} }
				}{\derLmu f{:}`w\prod `w\arrow `n,x{:}`w |- f(xx) : `w\arrow `n | {} }
			}{\derLmu f{:}`w\prod `w\arrow `n |- `lx.f(xx) : `w\prod `w\arrow `n | {} }
		 \kern-2cm
		 \Inf	[`w]
			{\derLmu f{:}`w\prod `w\arrow `n |- `lx.f(xx) : `w | {} }
		}{\derLmu f{:}`w\prod `w\arrow `n |- (`lx.f(xx))(`lx.f(xx)) : `w\arrow `n | {} }
	}{\derLmu |- `lf.(`lx.f(xx))(`lx.f(xx)) : (`w\prod `w\arrow `n)\prod `w\arrow `n | {} }
 \]
Notice that this term does not have a normal form, so is not strongly normalisable.

 \begin{definition} 
The set $ \SN$ of \emph{strongly normalisable} terms is defined as usual as the set of all terms $M$ such that no infinite reduction sequence out of $M$ exists; we use $\SN(M)$ for $M \ele \SN$, and $ \SN^*$ for the set of finite stacks of terms in $ \SN$.
 \end{definition}

The following is straightforward:

 \begin{proposition} \label {SN facts}
 \begin{enumerate}

 \item \label {SN fact head application}
If $\SN(x\Vec{M})$ and $\SN(\Vec{N})$, then $\SN(x\Vec{M}\Vec{N})$.

 \item \label {SN fact redex}
If $\SN(M[N/x]\Vec{P})$ and $\SN(N)$, then $\SN({( `lx.M)N\Vec{P}})$.

 \item \label{SN fact add mu abstraction}
If $\SN(M)$, then $\SN(`m`a.[`b]M)$.

 \item \label {SN fact mu in redex}
If $\SN( `m`a.[`b]M \strSubvec[ `a <= N ] \Vec{L} ) $ and $ \SN(\Vec{N}) $, then $ \SN({( `m`a.[`b]M )\Vec{N}\Vec{L}}) $.

 \item \label {SN fact mu out redex}
If $\SN(`m`a.[`a]M \strSubvec[ `a <= N ] \Vec{N} \Vec{L} ) $, then $ \SN({( `m`a.[`a]M )\Vec{N}\Vec{L}}) $.

 \end{enumerate}
 \end{proposition}

 \begin{definition}[Type Interpretation] \label{TypeSem definition}

 \begin{enumerate}
 \item
We define a map 
 \[ \begin{array}{rl}
 \TypeSem{\cdot}: & (\Lang_D \rightarrow \wp(\Terms)) + (\Lang_C \rightarrow \wp(\Terms^*) )
 \end{array} \] 
(where $\wp$ represents the powerset constructor) interpreting term types and continuation-stack types as, respectively, sets of terms and sets of stacks, as follows:
 \[ \begin{array}{rcccl}
 \TypeSem{`n} & = & \TypeSem{`w\arrow`n} &=& \SN 
	\\ &&
 \TypeSem{`k\arrow`n} &=& \Set{M \ele \Terms \mid \forall \Vec{L} \ele \TypeSem{`k} \Pred[ M \Vec{L} \ele \TypeSem{`n} ] } 
	\\ &&
 \TypeSem{`d\prod `w} &=& \Set{N \cons \Vec{L} \mid N \ele \TypeSem{`d}, \Vec{L} \ele \SN^*} 
	\\ &&
 \TypeSem{`d\prod `k} &=& \Set{N \cons \Vec{L} \mid N \ele \TypeSem{`d}, \Vec{L} \ele \TypeSem{`k}}
	\\ &&
 \TypeSem{`s \inter `t} &=& \TypeSem{`s} \cap \TypeSem{`t} \\
 \end{array} \]

 \item
We define the \emph{length} of a stack type, $ \length{`.} : \Lang_C \rightarrow \Nat$, as follows:
 \[ \begin{array}{ccl}
 \length{`d\prod `w} &=& 1 
 \\ 
 \length{`d\prod `k} &=& 1+ \length{`k} 
 \\ 
 \length{`k_1 \inters`k_2} &=&  \textit{max } \length{`k_1} ~ \length{`k_2} 
 \end{array} \]

 \end{enumerate}
 \end{definition}

By this interpretation, the elements of $ \TypeSem{`d_1\prod \dots \prod `d_n \prod `w}$ are
stacks of strongly normalisable terms that have an arbitrary length greater than or equal to $n$. 
It is easy to check that 
$ \length{`k}$ returns the minimal length of the stacks in $ \TypeSem{`k}$.

We can show:

 \begin{lemma} \label {TypeSem and SN lemma}
For any $ `d\ele\Lang_D$ and $`k\ele\Lang_C$:
 \begin{enumerate}

 \item \label{TypeSem implies SN}
$\TypeSem{`d} \subseteq \SN $ and $\TypeSem{`k} \subseteq \SN^* $.

 \item \label{SN head implies TypeSem}
$x\Vec{N} \ele \SN \Then x\Vec{N} \ele \TypeSem{`d}$.

 \item \label {vec x implies TypeSem}
$\Vec{x}=x_1\cons\ldots:x_n \ele \TypeSem{`k}$, for all $n$ such that $n\geq \length{`k}$ .

 \end{enumerate}

 \begin{proof}
By simultaneous induction on the structure of types, using Definition \ref {TypeSem definition}. %
{
We show some of the cases.
 \begin{enumerate} \itemsep 3pt

 \item 
 \begin {description}

\Comment{
 \item [$(`n, `w\arrow `n):$]
Immediate.
}

 \item [$(`k\arrow`n):$]
$ \begin{array}[t]{ll}
M \ele \TypeSem{`k\arrow`n} 
	& \Then (\IH (\ref {SN head implies TypeSem})) \\
 \Vec{x} \ele \TypeSem{`k} \And M \ele \TypeSem{`k\arrow`n} 
	& \Then (\ref {TypeSem definition}) \\
M\Vec{x} \ele \TypeSem{`n} 
	& \Then (\ref {TypeSem definition}) \\
M\Vec{x} \ele \SN 
	&\Then 
M \ele \SN.
 \end{array} $

\Comment{
 \item [$(`d_1\inters`d_2):$]
$ \begin{array}[t]{llllll}
M \ele \TypeSem{`d_1\inters`d_2} 
	&\Then& 
M \ele \TypeSem{`d_1} 
	&\Then (\IH (\ref {TypeSem implies SN})) & 
M \ele \SN .
 \end{array} $
}

 \item[$(`d\prod`w):$]
$ \begin{array}[t]{ll}
M \ele \TypeSem{`d\prod`w} 
	& \Then (\ref {TypeSem definition})) \\
M = N \cons \Vec{L} \And N \ele \TypeSem{`d} \And \Vec{L} \ele \SN 
	& \Then (\IH (\ref {TypeSem implies SN})) \\
N \ele \SN \And \Vec{L} \ele \SN^* 
	&\Then 
N \cons \Vec{L} \ele \SN^*.
 \end{array} $

 \item[$(`d\prod`k):$]
$ \begin{array}[t]{ll}
M \ele \TypeSem{`d\prod`k} 
	& \Then (\ref {TypeSem definition})) \\
M = N \cons \Vec{L} \And N \ele \TypeSem{`d} \And \Vec{L} \ele \TypeSem{`k} 
	& \Then (\IH (\ref {TypeSem implies SN})) \\
N \ele \SN \And \Vec{L} \ele \SN^* 
	&\Then 
N \cons \Vec{L} \ele \SN^*.
 \end{array} $

\Comment{
 \item [$(`k_1\inters`k_2):$]
$ \begin{array}[t]{llllll}
 \Vec{L} \ele \TypeSem{`k_1\inters`k_2} 
 	&\Then& 
\Vec{L} \ele \TypeSem{`k_1} 
	&\Then (\IH (\ref {TypeSem implies SN})) & 
\Vec{L} \ele \SN^*. 
 \end{array} $
}

 \end {description}

\Comment{
}

 \item
 \begin {description}

\Comment{
 \item [$(`n, `w\arrow `n):$]
Immediate.
}

 \item [$(`k\arrow`n):$]
$ \begin{array}[t]{lll}
 x\Vec{N} \ele \SN 
 	& \Then (\ref {TypeSem definition} \And \IH(\ref {TypeSem implies SN})) \\
 \Vec{L} \ele \TypeSem{`k} \Then x\Vec{N} \ele \SN \And \Vec{L} \ele \SN^* 
	& \Then (\ref {SN facts}) \\
 \Vec{L} \ele \TypeSem{`k} \Then x\Vec{N}\Vec{L} \ele \SN 
	& \Then (\IH(\ref {SN head implies TypeSem})) \\
 \Vec{L} \ele \TypeSem{`k} \Then x\Vec{N}\Vec{L} \ele \TypeSem{`n} 
	& \Then (\ref {TypeSem definition}) & 
x\Vec{N} \ele \TypeSem{`k\arrow`n}.
 \end{array}$

\Comment{
 \item [$(`d_1\inters`d_2):$]
$ \begin{array}[t]{llllll}
x\Vec{N} \ele \SN 
	& \Then (\IH(\ref{SN head implies TypeSem})) & 
x\Vec{N} \ele \TypeSem{`d_1} \And x\Vec{N} \ele \TypeSem{`d_2} 
	& \Then (\ref {TypeSem definition}) & 
x\Vec{N} \ele \TypeSem{`d_1\inters`d_2}.
 \end{array} $
}

 \end {description}

 \item
 \begin {description}

 \item[$(`d\prod`w):$]

 $ \begin{array}[t]{llll}
 \Vec{x} = x \cons \Vec{x}' 
	&\Then (\IH(\ref{SN head implies TypeSem})) \\
x \ele \TypeSem{`d} \And \Vec{x}' \ele \SN^* 
	&\Then (\ref{TypeSem definition})& 
 \Vec{x} \ele \TypeSem{`d\prod`w}. 
 \end{array} $

 \item[$(`d\prod`k):$]

 $ \begin{array}[t]{llll}
 \Vec{x} = x \cons \Vec{x}' 
	&\Then (\IH(\ref{SN head implies TypeSem}) \And \IH(\ref{vec x implies TypeSem})) \\
x \ele \TypeSem{`d} \And \Vec{x}' \ele \TypeSem{`k} 
	&\Then (\ref{TypeSem definition})& 
 \Vec{x} \ele \TypeSem{`d\prod`k}. 
 \end{array} $
\\ [-10pt] \hbox{~} \QED

\Comment{
 \item [$(`k_1\inters`k_2):$]
$ \begin{array}[t]{llllll}
 \Vec{x} \ele \SN^* 
	&\Then (\IH (\ref {vec x implies TypeSem})) & 
 \Vec{x} \ele \TypeSem{`k_1} \And \Vec{x} \ele \TypeSem{`k_2} 
	&\Then (\ref{TypeSem definition})& 
 \Vec{x} \ele \TypeSem{`k_1\inters`k_2}. 
 \end{array} $ \QED
}

 \end {description}
 \end {enumerate}

}
 \end{proof}
 \end {lemma}

The following result follows immediately from Lemma~\ref{TypeSem and SN lemma}\,(\ref{SN head implies TypeSem}):

 \begin{corollary} \label {variable implies D }
For any $x\ele\TVar$ and any $`d\ele\Lang_D$: $x \ele \TypeSem{`d}$.
 \end{corollary}

The following lemma shows that our type interpretation is closed under the type inclusion relation.

 \begin{lemma} \label{closed under leq}
For all $`s, `t \ele \Lang $: if $`s \leq `t$, then $\TypeSem{`s} \subseteq \TypeSem{\tau}$.

 \begin{proof}
By induction on the definition of $ \leq$. We show some of relevant cases.

 \begin{description}
 \Comment{
 \item[$(`s \inter `t \leq `s $, $`s \inter `t \leq `t ):$] 
Immediate by Definition~\ref{TypeSem definition}.

 \item[$(`s \leq `t _1 \And `s \leq `t _2 \Then `s \leq `t _1 \inter `t _2 ):$]
Immediate by induction and Definition~\ref{TypeSem definition}.


 \item[$(`d _1\prod `d _2\prod `w \leq `d _1\prod `w):$]
By Definition~\ref{TypeSem definition}, since $ \TypeSem{`d_2} \subseteq \SN$ by Lemma \ref{TypeSem and SN lemma}\,(\ref{TypeSem implies SN}).

}

 \myitem[$(`d _1\prod `w ) \inter(`d _2\prod `k ) \leq (`d _1 \inter`d _2)\prod `k ):$] 
 \TypeSem{(`d _1\prod `w ) \inter(`d _2\prod `k )}
	&=& \\
 \Set{M \cons \Vec{L} \mid M \ele \TypeSem{`d_1}, \Vec{L} \ele \SN^*} \cap
	\Set{M \cons \Vec{L} \mid M \ele \TypeSem{`d_2}, \Vec{L} \ele \TypeSem{`k}}
	&=& ( \textrm{$ \TypeSem{`k} \subseteq \SN^*$ by~\ref{TypeSem and SN lemma}\,(\ref{TypeSem implies SN}}) ) \\ 
 \Set{M \cons \Vec{L} \mid M \ele \TypeSem{`d_1}\cap \TypeSem{`d_2}, \Vec{L} \ele \TypeSem{`k}}
	&=& \\
 \Set{M \cons \Vec{L} \mid M \ele \TypeSem{`d_1\inters`d_2}, \Vec{L} \ele \TypeSem{`k}}
	&=& \\ 
 \TypeSem{(`d _1 \inter`d _2)\prod `k} 
 \end{array} $.

 \Comment{
 \item[$((`d _1\prod `k_1) \inter (`d _2\prod `k_2) \seq (`d _1 \inter`d _2)\prod (`k_1 \inter`k_2)$ if $`k_1,`k_2 \not \equiv`w ):$]\\
 \[ \begin{array}{lcll}
 \TypeSem{(`d _1\prod `k_1) \inter (`d _2\prod `k_2)}
	&=& \\
 \Set{M \cons \Vec{L} \mid M \ele \TypeSem{`d_1}, \Vec{L} \ele \TypeSem{`k_1}} \cap
	 \Set{M \cons \Vec{L} \mid M \ele \TypeSem{`d_2}, \Vec{L} \ele \TypeSem{`k_2}}
	&=& \\
 \Set{M \cons \Vec{L} \mid M \ele \TypeSem{`d _1} \cap \TypeSem{`d _2},
	 \Vec{L} \ele \TypeSem{`k_1} \cap \TypeSem{`k_2}}
	&=& \\
 \Set{M \cons \Vec{L} \mid M \ele \TypeSem{`d _1 \inter`d _2},
	 \Vec{L} \ele \TypeSem{`k_1 \inter`k_2}}
	&=& \TypeSem{(`d _1 \inter`d _2)\prod (`k_1 \inter`k_2)}
 \end{array} 
 \]

 \myitem[$( `d _1 \leq `d _2 \Then `d _1\prod `w \leq `d _2\prod `w ):$]
 \TypeSem{`d _1\prod `w }
	&=& \\
 \Set{M \cons \Vec{L} \mid M \ele \TypeSem{`d_1}, \Vec{L} \ele \SN^*}
	 & \subseteq & (\TypeSem{`d_1} \subseteq \TypeSem{`d_2}\textrm{by induction}) \\
 \Set{M \cons \Vec{L} \mid M \ele \TypeSem{`d_2}, ~ \Vec{L} \ele \SN^*}
	&=& \\
 \TypeSem{`d _2\prod `w }
 \end{array} $

 \myitem[$( `d_1 \seq `d _2 \And `k_1 \seq `k_2 \Then `d _1\prod `k_1 \seq `d _2\prod `k_2 ):$]
 \TypeSem{`d _1\prod `k_1 }
	&=& \\
 \Set{M \cons \Vec{L} \mid M \ele \TypeSem{`d_1}, \Vec{L} \ele \TypeSem{`k_1}}
	& \subseteq & (\TypeSem{`d_1} \subseteq \TypeSem{`d_2}\textrm{ and } \TypeSem{`k_1} \subseteq \TypeSem{`k_2} \textrm{by induction}) \\
 \Set{M \cons \Vec{L} \mid M \ele \TypeSem{`d_2}, \Vec{L} \ele \TypeSem{`k_2}}
	&=& \\
 \TypeSem{`d _2\prod `k_2}
 \end{array} $
}

 \myitem[$( `k_2 \leq `k_1 \Then `k_1\arrow`n \leq `k_2\arrow`n ):$]
 \TypeSem{`k_1\arrow`n}
	&=& \\
 \Set{M \ele \Terms \mid \forall \Vec{L} \ele \TypeSem{`k_1} \Pred[ M \Vec{L} \ele \SN ] }
	& \subseteq & (\TypeSem{`k_2} \subseteq \TypeSem{`k_1}\textrm{ by induction} ) \\
 \Set{M \ele \Terms \mid \forall \Vec{L} \ele \TypeSem{`k_2} \Pred[ M \Vec{L} \ele \SN ] }
	&=& \\
 \TypeSem{`k_2\arrow`n} \\
 \end{array} $ 
\\ [-10pt] \hbox{~} \QED

 \Comment{
 \item[$(`n = `w\arrow`n):$]
$ \TypeSem{`n} = \SN = \TypeSem{`w\arrow`n}$.
}

 \end{description}
 \end{proof}
 \end{lemma}

Our type interpretation is closed under expansion for the logical and for the structural reduction, with the proviso that the term or stack to be substituted is an element of an interpreted type as well.

 \begin{lemma} \label {Typesem expansion} \label{lem:pluriSat}
For any $`d,`d'\ele\Lang_D$ and $`k\ele\Lang_C$:
 \begin{enumerate}

 \item \label {Typesem expansion logical}
If $ M[N/x]\Vec{P} \ele \TSem{`d} $ and $ N \ele \TSem{`d'} $, then $ ( `lx.M)N\Vec{P} \ele \TSem{`d}$.

 \item \label {Typesem expansion structural in}
If $ `m`a.[`b]M \strSubvec [ `a <= N ] \Vec{P} \ele \TSem{`d} $ and $\Vec{N} \ele \TSem{`k}$, then $ (`m`a.[`b]M) \Vec{N} \Vec{P} \ele \TSem{`d} $.

 \item \label {Typesem expansion structural out}
If $ `m`a.[`a]M \strSubvec [ `a <= N ] \Vec{N} \Vec{P} \ele \TSem{`d} $, then $ (`m`a.[`a]M) \Vec{N} \Vec{P} \ele \TSem{`d} $.

 \end{enumerate}

 \begin{proof}
By induction on the structure of types, using \ref {SN facts}, \ref {TypeSem definition} and \ref{TypeSem and SN lemma}.\QED

 \Comment{
 \begin{enumerate}

 \item
By induction on the structure of types.

 \begin {description}

 \item [$(`s = `n, ~ `w\arrow`n):$]
$ \begin {array}[t]{ll}
M[N/x]\Vec{P} \ele \TSem{`s} \And N \ele \TSem{`d'}
	& \Then (\ref {TypeSem definition} \And \ref{TypeSem and SN lemma} ) \\
 \SN(M[N/x]\Vec{P}) \And \SN(N)
	& \Then ({\ref {SN facts}\,(\ref {SN fact redex}) }) \\
 \SN({( `lx.M)N\Vec{P}})
	& \Then (\ref {TypeSem definition} ) \\
( `lx.M)N\Vec{P} \ele \TSem{`s}
 \end {array} $

 \item [$(`k\arrow`n):$]
$ \begin {array}[t]{ll}
M[N/x]\Vec{P} \ele \TSem{`k\arrow`n} \And N \ele \TSem{`d'}
	& \Then (\ref {TypeSem definition}) \\
 \Vec{Q} \ele \TSem{`k} \Then M[N/x]\Vec{P}\Vec{Q} \ele \TSem{`n} \And N \ele \TSem{`d'}
	& \Then (\IH) \\
 \Vec{Q} \ele \TSem{`k} \Then (`lx.M)N\Vec{P}\Vec{Q} \ele \TSem{`n}
	& \Then (\ref {TypeSem definition}) \\
(`lx.M)N\Vec{P} \ele \TSem{`k\arrow`n}
 \end {array} $

 \item [$(`d_1\inters`d_2):$]
Directly by induction.

 \end {description}

 \item
By induction on the structure of types.

 \begin{description}

 \item [$(`s = `n, ~ `w\arrow`n):$]
$ \begin {array}[t]{llll}
`m`a.[`b]M \strSubvec [ `a <= N ] \Vec{P} \ele \TSem{`s} \And \Vec{N} \ele \TSem{`k} 
	& \Then (\ref {TypeSem definition} \And \ref{TypeSem and SN lemma} ) \\
 \SN(`m`a.[`b]M \strSubvec[ `a <= N ]\Vec{P}) \And \SN(\Vec{N}) 
	& \Then ({\ref {SN facts}\,(\ref {SN fact mu in redex}) }) \\
 \SN({( `m`a.[`b]M)\Vec{N}\Vec{P}}) 
	& \Then (\ref {TypeSem definition} ) \\
( `m`a.[`b]M)\Vec{N}\Vec{P} \ele \TSem{`s}
 \end {array} $

 \item [$(`k'\arrow`n):$]
$ \begin {array}[t]{llll}
`m`a.[`b]M \strSubvec [ `a <= N ] \Vec{P} \ele \TSem{`k'\arrow`n} \And \Vec{N} \ele \TSem{`k}
	& \Then (\ref {TypeSem definition}) \\
 \Vec{Q} \ele \TSem{`k'} \Then `m`a.[`b]M \strSubvec [ `a <= N ] \Vec{P} \Vec{Q} \ele \TSem{`n} \And \Vec{N} \ele \TSem{`k}
	& \Then (\IH) \\
 \Vec{Q} \ele \TSem{`k'} \Then (`m`a.[`b]M) \Vec{N} \Vec{P} \Vec{Q} \ele \TSem{`n}
	& \Then (\ref {TypeSem definition}) \\
(`m`a.[`b]M) \Vec{N} \Vec{P} \ele \TSem{`k'\arrow`n}
 \end {array} $

 \item [$(`d_1\inters`d_2):$]
Directly by induction.

 \end{description}

 \item
By induction on the structure of types.

 \begin{description}

 \item [$(`s = `n, ~ `w\arrow`n):$]
$ \begin {array}[t]{ll}
`m`a.[`a]M \strSubvec [ `a <= N ] \Vec{P} \ele \TSem{`s} 
	& \Then (\ref {TypeSem definition} \And \ref{TypeSem and SN lemma} ) \\
 \SN(`m`a.[`a]M \strSubvec[ `a <= N ]\Vec{P}) 
 	& \Then ({\ref {SN facts}\,(\ref {SN fact mu out redex}) }) \\
 \SN({( `m`a.[`a]M)\Vec{N}\Vec{P}}) 
 	& \Then (\ref {TypeSem definition} ) \\
( `m`a.[`a]M)\Vec{N}\Vec{P} \ele \TSem{`s}
 \end {array} $

 \item [$(`k'\arrow`n):$]
$ \begin {array}[t]{ll}
`m`a.[`a]M \strSubvec [ `a <= N ] \Vec{P} \ele \TSem{`k'\arrow`n}
	& \Then (\ref {TypeSem definition}) \\
 \Vec{Q} \ele \TSem{`k'} \Then `m`a.[`a]M \strSubvec [ `a <= N ] \Vec{P} \Vec{Q} \ele \TSem{`n}
	& \Then (\IH) \\
 \Vec{Q} \ele \TSem{`k'} \Then (`m`a.[`a]M) \Vec{N} \Vec{P} \Vec{Q} \ele \TSem{`n}
	& \Then (\ref {TypeSem definition}) \\
(`m`a.[`a]M) \Vec{N} \Vec{P} \ele \TSem{`k'\arrow`n}
 \end {array} $

 \item [$(`d_1\inters`d_2):$]
Directly by induction.

 \end{description}
 \end{enumerate}
}\Comment

 \end{proof}
 \end{lemma}

In Theorem~\ref{thr:typableAreSN} we will show that all typeable terms are strongly normalisable.
In order to achieve that, we first show, in Lemma~\ref{lem:replacement}, that for any a term $M$ typeable with $`d$, any full substitution instance $M_{\Repl}$  (i.e.~replacing all free term variables by terms, and feeding stacks to all free names) is an element of the interpretation of $`d$, which by Lemma~\ref{TypeSem and SN lemma} implies that $M_{\Repl}$ is strongly normalisable.
We need these substitutions to be applied all `in one go', so define a notion of parallel substitution.
The main result is then obtained by taking the substitution that replaces term variables by themselves and names by stacks of term variables.
The reason we first prove the result for \emph{any} substitution is that, in the proof of  
Lemma~\ref{lem:replacement}, in the case for $`lx.M$ and $`m`a.Q$ the substitution is extended, by replacing the bound variable or name with a normal term (or stack).

 \begin{definition}
 \begin{enumerate}

 \item
A partial mapping $ \Repl : (\TVar\rightarrow\Terms) + (\CVar\rightarrow\Terms^*)$ is a \emph{parallel substitution} if, for every $\textsl{p},\textsl{q} \ele \dom (\Repl) $, if $\textsl{p}\not=\textsl{q}$ then $\textsl{p} \notele \fv(\Repl \textsl{q})$ and $\textsl{p} \notele \fn(\Repl \textsl{q})$.

 \item
Borrowing a notation for valuations, for a parallel substitution $\Repl$ we define the application of $\Repl$ to a term by:
 \[ \begin{array}{rcll}
([`a]M)_{\Repl}  &\ByDef& [`a] M_{\Repl} \Vec{L} & \textit{if } \Repl `a = \Vec{L} 
	\\
([`b]M)_{\Repl}  &\ByDef& [`b]M_{\Repl} & \textit{if } `b \notele \dom(\Repl) 
	\\
(`m`b .Q)_{\Repl}  &\ByDef & `m`b.Q_{\Repl}  
\\ 
x_{\Repl}  & \ByDef & N & \textit{if } \Repl x = N 
	\\
y_{\Repl} & \ByDef & y & \textit{if } y \notele \dom(\Repl) 
	\\
(`l x.M)_{\Repl} & \ByDef & `l x.M_{\Repl} 
	\\
(MN)_{\Repl} & \ByDef & M_{\Repl} N_{\Repl} 
 \end{array} \]

 \item
We define $ \Repl [ N/x ]$ and $ \Repl \strSubvec[ `a <= L ] $ by, respectively,
 \[ \begin{array}{rcl@{\qquad}rcl}
 \Repl [ N/x ] \, y &\ByDef& 
 \begin{cases}{ll}
N & \textrm{if } y=x \\
 \Repl \, y & \textrm{otherwise}
 \end{cases}
\\ [5mm]
 \Repl \strSubvec[ `a <= L ] \, `b &\ByDef& 
 \begin{cases}{ll}
 \Vec{L} & \textrm{if } `a=`b \\
 \Repl \, `b & \textrm{otherwise}
 \end{cases}
 \end{array} \]
 
 \item
We will say that $\Repl \mbox{ \emph{extends} } `G \mbox{ \emph{and} } `D$, if, for all $x{:}`d \ele `G$
and $`a{:}`k \ele `D$, we have, respectively, $\Repl (x) \ele \TypeSem{`d}$ and $\Repl (`a) \ele \TypeSem{`k}$.


 \end{enumerate}
 \end{definition}

Notice that we do allow a variable to appear in its own image under $\Repl$.
Since $x$ does not appear in $M[N/x]$, this does not violate Barendregt's convention.

 \begin{lemma}[Replacement Lemma] \label{lem:replacement}
Let $ \Repl $ be a parallel substitution that extends $`G$ and $`D$.
Then:
 \[ \begin{array}{rcl}
 \textit{if }~ \derLmu `G |- M : `d | `D &\textit{then}& M_{\Repl} \ele \TypeSem{`d}. 
 \end{array} \]

 \begin{proof}
By induction on the structure of derivations. 
We show some more illustrative cases.

 \begin{description}

 \Comment{
 \item[$(\Ax):$] then $M = x$.
Since $x{:}`d \in`G$ and $\Repl $ extends $`G$, so $M_{\Repl} = \Repl (x) \ele \TypeSem{`d}$.
}

 \item [$(\Abs):$] Then $M = `lx.M'$, $`d = `d'\prod `k\arrow`n $, and $ \derLmu `G,x{:}`d' |- M' : `k\arrow`n | `D $.
Take $N \ele \TypeSem{`d'}$; since $x$ is bound, by Barendregt's convention we can assume that it does not occur free in the image of $ \Repl $, so $ \Repl [ N/x ] $ is a well-defined parallel substitution that extends $`G, x{:}`d'$ and $`D$.
Then by induction, we have $M'{\indexRepl [ N/x ]} \ele \TypeSem{`k\arrow`n}$.
Since $x$ does not occur free in the image of $ \Repl $, $M'{\indexRepl [ N/x ] } = M'{\indexRepl}[N/x]$, so also $M'{\indexRepl}[N/x] \ele \TypeSem{`k\arrow`n}$.
By Lemma~\ref{Typesem expansion}\,(\ref{Typesem expansion logical}), also
$(`lx.M'{\indexRepl})N \ele \TypeSem{`k\arrow`n}$.
By definition of $ \TypeSem{`k\arrow`n}$, for any $ \Vec{L} \ele \TypeSem{`k}$ we have $(`lx.M'{\indexRepl} )N \Vec{L} \ele \TypeSem{`n}$; notice that $N \cons \Vec{L} \ele \TypeSem{`d \prod `k}$, so
$(`lx.M'){\indexRepl} \ele \TypeSem{`d'\prod `k\arrow`n}$.

 \Comment{
 \item[$(\App):$]
Then $M = PQ$, and there exists $`d$ such that $ \derLmu `G |- P : `d\prod `k\arrow`n | `D $ and $ \derLmu `G |- Q : `d | `D $.
Then by induction, we have $P{\indexRepl} \ele \TypeSem{`d\prod `k\arrow`n}$ and $Q{\indexRepl} \ele \TypeSem{`d}$; notice that $P{\indexRepl}Q{\indexRepl} = (PQ){\indexRepl}$.
We now distinguish two cases:

 \begin{description}

 \item[$(`k \not= `w ):$]
Take $ \Vec{L} \ele \TypeSem{`k}$, we get $Q{\indexRepl} \cons \Vec{L} \ele \TypeSem{`d\prod `k}$, so also $P{\indexRepl}Q{\indexRepl} \Vec{L} \ele \TypeSem{`n}$.
But then also $P{\indexRepl}Q{\indexRepl} \ele \TypeSem{`k\arrow `n}$.

 \item[$(`k = `w ):$]
We need to show that $P{\indexRepl}Q{\indexRepl} \ele \TypeSem{`w\arrow `n}$, for which it suffices to show that $P{\indexRepl}Q{\indexRepl} \ele \SN$.
Take $ \Vec{L} \ele \SN^*$, then $Q{\indexRepl} \cons \Vec{L} \ele \TypeSem{`d\prod `w}$, so $P{\indexRepl}Q{\indexRepl} \Vec{L} \ele \TypeSem{`n} $, so $P{\indexRepl}R{\indexRepl} \Vec{L} \ele \SN $; in particular $(PQ){\indexRepl} \ele \SN = \TypeSem{`w\arrow`n}$.

 \end{description}

}

 \item[$(`m):$] 
Then $M = `m`a.[`b]M'$, and $`d = `k \arrow `n $.
We distinguish two different sub-cases.

 
 \item[\quad $`a = `b:$]
Then $M = `m`a. [`a]M'$, $`d = `k \arrow `n$, and $ \derLmu `G |- M' : `k\arrow`n | `a{:}`k,`D $.
Take $ \Vec{L} \ele \TypeSem{`k}$; since $`a$ is bound in $M$, we can assume it does not occur free in the image of $\Repl$, so $ \Repl \strSubvec[ `a <= L ] $ is a well-defined parallel substitution that extends $`G$ and $`D,`a{:} `k$, and by induction, $M'_{\Repl \scstrSubvec[ `a <= L ]  } \ele \TypeSem{`k\arrow`n}$.
Since $`a$ does not occur free in the image of $\Repl$, $M'_{\Repl\scstrSubvec[ `a <= L ]  } = M'{\indexRepl} \strSubvec[ `a <= L ]$, so we have $M'{\indexRepl}\strSubvec[ `a <= L ]  \ele \TypeSem{`k\arrow`n} $, and therefore $M'{\indexRepl} \strSubvec[ `a <= L ] \Vec{L} \ele \TypeSem{`n}$.
Then by Definition~\ref{TypeSem definition}, $\SN(M'{\indexRepl} \strSubvec[ `a <= L ] \Vec{L})$, but then also ${\SN(`m`a.[`a]M'{\indexRepl} \strSubvec[ `a <= L ] \Vec{L})}$, by Lemma~\ref{SN facts}\,(\ref{SN fact add mu abstraction}). 
So $ `m`a.[`a]M'{\indexRepl} \strSubvec[ `a <= L ] \Vec{L} \ele \TypeSem{`n} $. 
Then by Lemma~\ref{Typesem expansion}\,(\ref{Typesem expansion structural out}), $(`m`a.[`a]M'{\indexRepl}) \Vec{L} \ele \TypeSem{`n}$; so $ (`m`a.[`a]M'){\indexRepl} \ele \TypeSem{`k\arrow`n}$.

 \item[\quad $`a \not= `b:$]
Then $`D = `b{:}`k',`D' $, and $ \derLmu `G |- M' : `k'\arrow`n | `a{:}`k,`b{:}`k',`D $.
Assume $\Vec{L} \ele \TypeSem{`k}$, then $\Repl\strSubvec[ `a <= L ]  $ extends $`G$ and $`a{:}`k, `b{:}`k',`D' $.
Then, by induction, $M'_{\Repl \scstrSubvec[ `a <= L ]  } \ele \TypeSem{`k' \arrow`n} $.
Now let $\Vec{Q} \ele \TypeSem{`k'}$, then $ M'_{\Repl \scstrSubvec[ `a <= L ] } \Vec{Q} \ele \TypeSem{`n} $ and then also $ (M' \Vec{Q} )_{\Repl \scstrSubvec[ `a <= L ] } \ele \TypeSem{`n} $.\\
Then $\SN((M' \Vec{Q} )_{\Repl \scstrSubvec[ `a <= L ] })$ by Definition~\ref{TypeSem definition}, and $\SN(`m`a.[`b](M' \Vec{Q} )_{\Repl \scstrSubvec[ `a <= L ] })$  by Lemma \ref{SN facts}\,(\ref{SN fact add mu abstraction}), so, again by Definition~\ref{TypeSem definition},  $`m`a.[`b](M' \Vec{Q} )_{\Repl \scstrSubvec[ `a <= L ] } \ele \TypeSem{`n} $.
As in the previous part, $`a$ is not free in the image of $\Repl$,
and therefore also $ `m`a.[`b](M' \Vec{Q} ){\indexRepl} {\strSubvec[ `a <= L ] } \ele \TypeSem{`n} $.

Then, by Lemma~\ref{Typesem expansion}\,(\ref{Typesem expansion structural in}), $ (`m`a.[`b](M'\Vec{Q}){\indexRepl})\Vec{L} \ele \TypeSem{`n} $.
Notice that $[`b]M'{\indexRepl} \Vec{Q} = [`b]M'{\indexRepl} \strSubvec[ `b <= Q ] $; since $\Repl{`b} = \Vec{Q}$, we can infer that $[`b]M'{\indexRepl} \Vec{Q} = [`b]M'{\indexRepl} $, so $ (`m`a.[`b]M'){\indexRepl} \Vec{L} \ele \TypeSem{`n} $.
But then $ (`m`a.[`b]M'){\indexRepl} \ele \TypeSem{`k\arrow`n} $.
 \QED 


 \Comment{

 \item[$(\inters):$]
By induction, using the definition of Type Interpretation.

 \item[$(\leq):$]
By induction and Lemma~\ref{closed under leq}.
 
}

 \end{description}
 \end{proof}
 \end{lemma}

We now come to the main result of this section, that states that all terms typeable in our system are strongly normalisable.

 \begin{theorem}[Typeable terms are $ \SN$] \label{thr:typableAreSN}
If $\derLmu `G |- M : `d | `D $ for some $`G$, $`D$ and $`d$, then $M \ele \SN$.

 \begin{proof}
Let $ \Repl $ be a parallel substitution such that
 \[ \begin{array}{rcl@{\dquad}l}
 \Repl (x) &=& x & \textrm{for } x \ele \dom(`G) \\
 \Repl (`a) &=& \Vec{y}_{`a} & \textrm{for } `a \ele \dom(`D) 
 \end{array} \]
where the length of the stack $ \Vec{y}_{`a}$ is $\length{`k}$ if $`a{:}`k \ele `D$
(notice that $\Repl$ is well defined).
By Lemma~\ref{TypeSem and SN lemma}, $ \Repl $ extends $`G$ and $`D$.
Hence, by Lemma~\ref{lem:replacement}, $M{\indexRepl} \ele \TypeSem{`d}$,
and then $M{\indexRepl} \ele \SN$ by Lemma~\ref{TypeSem and SN lemma}\,(\ref{TypeSem implies SN}).
Now 
 \[ \begin{array}{rcl}
M_{\Repl} 
	&\equiv& 
M \,[ x_1/ x_1, \ldots, x_n/x_n, \StrSubNoB{`a_1}{\Vec{y}_{`a_1}}, \ldots, \StrSubNoB{`a_m}{\Vec{y}_{`a_m}}] \\
	&\equiv& 
M \,[ \StrSubNoB{`a_1}{\Vec{y}_{`a_1}}, \ldots, \StrSubNoB{`a_m}{\Vec{y}_{`a_m}}] 
 \end{array} \]
Then, by Proposition~\ref{SN facts}, for any $\Vec{`b}$ also $(`m`a_1.[`b_1]\dots `m`a_m.[`b_m]M) \Vec{y}_{`a_1} \dots \Vec{y}_{`a_m} \ele \SN$, and therefore also $M \ele \SN$ .
 \QED

 \end{proof}
 \end{theorem}

 \subsection{Strongly Normalising Terms are Typeable} \label{sub:NormType}

In this section we will show the counterpart of the previous result, namely that all strongly normalisable terms are typeable in our intersection system.
This result has been claimed in many papers \cite{Pottinger'80,Bakel-TCS'92}, but has rarely been proven completely.

First we describe the shape of the terms in normal form. 

 \begin{definition}[Normal Forms] \label{def:normalForms}
The set $ \NF \subseteq \Terms$ of \emph{normal forms} is defined by the grammar:
 \[ \begin{array}{rcl}
N &::=& x N_1 \cdots N_k \mid `lx.N \mid `m`a.[`b]N 
 \end{array} \]
 \end{definition}
%
%
%
It is straightforward to verify that the terms in $\NF$ are precisely the irreducible ones.\\

We can show that all terms in $\NF$ are typeable.

 \begin{lemma} \label{prop:NFareTypable} \label{from normal form}
If $N \ele \NF$ then there exist $`G$, $`D$, and a type $`k\arrow `n$ such that $ \derLmu `G |- N : `k\arrow`n | `D $.

 \begin{proof}
By induction on the definition of $\NF$. We show the most relevant cases.
 \begin{description}

 \item [$(N \equiv xN_1 \ldots N_k):$] 
Since $N_1, \ldots,N_k \ele \NF$, by induction we have that, for all $i \leq k$ there exist $`G_i$, $`D_i$ and $`d_i$ such that $ \derLmu `G_i |- N_i : `d_i | `D _i $ (the structure of each $`d_i$ plays no role in this part). 
Take \[ `G = `G_1 \inters \dots \inters `G_k \inters x{:}(`d_1 \prod \cdots \prod `d_k \prod `d \prod `w) \arrow`n ,\textrm{ and } `D = `D_1 \inters \dots \inters `D_k .
 \]
where $`d$ is any element of $\Lang_D$. 
Then, by Lemma~\ref{restricted weakening}, $ \derLmu `G |- N_i : `d_i | `D $ for all $i \leq n$, and 
$ \derLmu `G |- x : {(`d_1 \prod \cdots \prod `d_k \prod `d \prod `w) \arrow`n} | `D $.
By repeated application of $(\App)$ we get $ \derLmu `G |- xN_1\dots N_k : `k\arrow`n | `D $
for $`k=`d\prod`w$.
 \Comment{
 \item [$(N \equiv `lx.N'):$]
By induction, there exist $`G,x{:}`d'$ and $`D$ such that $\derLmu `G,x{:}`d' |- N' : `k\arrow `n | `D $.
Then by $(\Abs)$ we obtain $\derLmu `G |- `lx.N' : `d\prod`k\arrow `n | `D $.
}

 \item [{$(N \equiv `m`a.[`b]N'):$}]
By induction, $ \derLmu`G |- N' : `k\arrow`n | `D $. 
We distinguish two cases:

 \begin{description}
 \item[$(`a \equiv`b):$]
In case $`a \ele \fn(N')$ and $`D=`a{:}`k',`D'$, 
we can construct:
 \[
\Inf	[`m]
	{\Inf	[\leq]
		{\Inf	[\Weak]
			{\InfBox{ \derLmu`G |- N' : `k\arrow`n | `a{:}`k',`D' }
			}{ \derLmu`G |- N' : `k\arrow`n | `a{:}`k\inters`k',`D' }
		}{ \derLmu`G |- N' : `k\inter`k'\arrow`n | `a{:}`k\inters`k',`D' }
	}{ \derLmu`G |- `m`a.[`a]N' : `k\inter`k'\arrow`n | `D' }
 \]
In case $`a \notele \fn(N')$, we can construct
 \[
\Inf	[`m]
	{\Inf	[\Weak]
		{\InfBox{ \derLmu`G |- N' : `k\arrow`n | `D }
		}{ \derLmu`G |- N' : `k\arrow`n | `a{:}`k,`D' }
	}{ \derLmu`G |- `m`a.[`a]N' : `k\arrow`n | `D' }
 \]

 \item[$(`a \not \equiv`b):$] 
We can proceed as in the previous case, obtaining now 
$ \derLmu`G |- N : `k\arrow`n | `a{:}`k',`b{:}`k,`D' $.
So by rule $(`m)$ we get $ \derLmu`G |- `m`a.[`b]N' : `k'\arrow`n | `b{:}`k'',`D' $.\QED
 
 \end{description}
 \end{description}
 \end{proof}
 \end{lemma}

We will now show that typeing is closed under expansion with respect to both logical and structural reduction, with the proviso that the term (stack) that gets substituted is typeable as well in the same contexts.

 \begin{lemma}[Contractum Expansion] \label{lem:redexExpBeta}\label{lem:redexExpMu} 
 \begin{enumerate}
 \item \label{lem:redexExpMu-i} 
If $ \derLmu `G |- M[N/x] : `d | `D $ and $ \derLmu `G |- N : `d' | `D $ then \\ $ \derLmu `G |- (`lx.M)N : `d | `D $.

 \item\label{lem:redexExpMu-ii}
If $ \derLmu `G |- `m`a . [`b]M \strSub[`a <= N ] : `d | `D $ and $ \derLmu `G |- N : `d' | `D $ then $ \derLmu `G |- (`m`a.[`b]M)N : `d | `D $.

 \end{enumerate}

 \begin{proof}

 \begin{enumerate}

 \item
Much the same as the similar result for the intersection systems for the $`l$-calculus.

 \Comment{

We consider two cases:
 \begin{description}
 
 \item [$(x \notele \fv(M)):$] 
Then $M[N/x] \equiv M$ and we may assume $x \notele \Dom(`G)$; by Lemma~\ref{lem:weakening} we have $ \derLmu `G,x{:}`d' |- M : `d | `D $ and we obtain $ \derLmu `G |- (`lx.M)N : `d | `D $ by using rule $(@)$ of Lemma~\ref{lem:EEadmissible}.

 \item[$(x \ele \fv(M)):$]
Since there exists a derivation that justifies $ \derLmu `G |- M[N/x] : `d | `D $, and $N$ is a true sub-term of $M[N/x]$, there exists $k\geq 1$ sub-derivations that have a judgement for $N$ in their conclusion, i.e.~such that there exists $`d_1, \ldots,`D_k$ such that $ \derLmu `G |- N : `d_i | `D $ for all $i \leq k$.
Then we can derive $ \derLmu `G,x{:}`d_1\inters \dots \inters `d_k |- M : `d | `D $ by replacing, for all $j \leq k$, each sub-derivation for $ \derLmu `G |- N : `d_j | `D $ by:
 \[
 \Inf	[ `d_1\inters \dots \inters `d_k \leq `d_j]
	{\derLmu `G,x{:}`d_1\inters \dots \inters `d_k |- x : `d_1\inters \dots \inters `d_k | `D
	}{\derLmu `G,x{:}`d_1\inters \dots \inters `d_k |- x : `d_j | `D }
 \]
By $(\intersI)$, we have $ \derLmu `G |- N : `d_1\inters \dots \inters `d_k | `D $, and we obtain $ \derLmu `G |- (`lx.M)N : `d | `D $ by using rule $(@)$ of Lemma~\ref{lem:EEadmissible}.

 \end{description}
}

 \item
We need to consider two different cases:
 \begin{description}
 
 \item[{$(`a \notele \fn([`b]M)):$}] 
Then $([`b]M) \strSub[`a <= N] \equiv [`b]M$ and $`a \not \equiv`b$.
We consider all the $n$ minimal sub-derivations ($n\geq 1$) having $`m`a.[`b]M$ as subject, from which conclusions we derive $ \derLmu `G |- `m`a . [`b]M \strSub[`a <= N ] : `d | `D $ by applying any number of $(\leq)$ and $(\inters)$ rules. 

The last step in each of these derivations is of the shape:
 \[ \begin{array}{c}
 \Inf	[`m]
	{\InfBox{ \derLmu `G |- M : `k\arrow`n | `a{:}`k_i,`b{:}`k,`D' }
	}{\derLmu `G |- `m`a.[`b]M : `k_i\arrow`n | `b{:}`k,`D' }
 \end{array} \]
where $`D = `b{:}`k,`D'$. 
Since $`a \notele \fn([`b]M)$, by strengthening (Lemma~\ref{restricted strengthening}) we can remove $`a{:}`k_i$ from the name context, so also $\derLmu `G |- M : `k\arrow`n | `b{:}`k,`D' $; then, by weakening (Lemma~\ref{lem:weakening}), we can add $`a{:}`d'\prod `k_i,$, so $ \derLmu `G |- M : `k\arrow`n | `a{:}`d'\prod `k_i,`b{:}`k,`D' $,
and then we can construct
 \[ \begin{array}{c}
 \Inf	[\App]
	{\Inf	[`m]
		{\InfBox{ \derLmu `G |- M : `k\arrow`n | `a{:}`d'\prod `k_i,`b{:}`k,`D' }
		}{\derLmu `G |- `m`a.[`b]M : `d'\prod `k_i\arrow`n | `b{:}`k,`D' }
	 \quad
	 \InfBox{ \derLmu `G |- N : `d' | `D }
	}{\derLmu `G |- (`m`a.[`b]M)N : `k_i\arrow`n | `D }
 \end{array} \]
from which it is possible to derive $ \derLmu `G |- (`m`a.[`b]M)N : `d | `D $
by applying the same $(\leq)$ and $(\inters)$ rules mentioned above.


 \item[{$(`a \ele \fn([`b]M)):$}]

We distinguish two further cases:

 \Comment{

 \[
 \Inf	{\derLmu `G |- M : `k'\arrow`n | `a{:}`k,`b{:}`k',`D
	}{\derLmu `G |- `m`a.[`b]M : `k\arrow`n | `b{:}`k ',`D }
 \quad
 \Inf	{\derLmu `G |- M : `k\arrow`n | `a{:}`k,`D
	}{\derLmu `G |- `m`a. [`a]M : `k\arrow`n | `D }
 \]
}

 \item[\quad $(`a =`b):$]
Then $([`b]M) \strSub[`a <= N] \equiv ([`a]M) \strSub[`a <= N] 
 \equiv [`a](M \strSub[`a <= N])N$;
we can assume, without loss of generality, that $`d = (`k_1\arrow`n)\inters \dots \inters (`k_n\arrow`n) $, and that for all $i \leq n$ there are sub-derivations constructed like
 \[
 \Inf	[`m]
	{\Inf	[\App]
		{\InfBox{ \derLmu `G |- M \strSub[`a <= N] : `d_i\prod `k_i\arrow`n | `a{:}`k_i,`D }
		 \dquad
		 \InfBox{ \derLmu `G |- N : `d_i | `D }
		}{\derLmu `G |- (M \strSub[`a <= N])N : `k_i\arrow`n | `a{:}`k_i,`D }
	}{\derLmu `G |- `m`a.[`a](M \strSub[`a <= N])N : `k_i\arrow`n | `D }
 \]

Then there exists $`d'_i$ such that $ \derLmu `G |- N : `d'_i | `D $, 
and $ \derLmu `G |- M : `d_i\prod `k_i\arrow`n | `a{:}`d'_i\prod `k_i,`D $ by Lemma \ref{lem:substitution}\,(\ref{lem:substitution ii}); so we can build the derivation:
 \[
 \Inf	[\App]
	{\Inf	[`m]
		{\Inf	[\Weak]
			{\Inf	[\leq]
				{\InfBox{ \derLmu `G |- M : `d_i\prod `k_i\arrow`n | `a{:}`d'_i\prod `k_i,`D }
				}{\derLmu `G |- M : `d_i\inters`d'_i\prod `k_i\arrow`n | `a{:}`d'_i\prod `k_i,`D }
			}{\derLmu `G |- M : `d_i\inters`d'_i\prod `k_i\arrow`n | `a{:}`d_i\inters`d'_i\prod `k_i,`D }
		}{\derLmu `G |- `m`a.[`a]M : `d_i\inters`d'_i\prod `k_i\arrow`n | `D }
	 \Inf	[\intersI]
	 	{\InfBox{ \derLmu `G |- N : `d_i | `D }
		 \dquad
		 \InfBox{ \derLmu `G |- N : `d'_i | `D }
	 	}{\derLmu `G |- N : `d_i\inters`d'_i | `D }
	}{\derLmu `G |- (`m`a.[`a]M)N : `k_i\arrow`n | `D }
 \]
We derive $ \derLmu `G |- (`m`a.[`a]M)N : `d | `D $ by rule $(\inters)$. 

 \item[\quad $(`a \not=`b):$]
Then $([`b]M) \strSub[`a <= N] \equiv [`b](M \strSub[`a <= N])$;
as above $`d = (`k_1\arrow`n)\inters \dots \inters (`k_n\arrow`n) $, and for all $i \leq n$ there are derivations structured like:
 \[
 \Inf	[`m]
	{\InfBox{ \derLmu `G |- M \strSub[`a <= N] : `k_i'\arrow`n | `a{:}`k_i,`b{:}`k_i',`D' }
	}{\derLmu `G |- `m`a.[`b](M \strSub[`a <= N]) : `k_i\arrow`n | `b{:}`k_i',`D' }
 \]
where $`D = `b{:}`k_i',`D'$.
As above, by Lemma~\ref{lem:substitution}\,(\ref{lem:substitution ii}) there exists $`d_i$ such that both \\ $ \derLmu `G |- N : `d_i | `b{:}`k_i',`D' $~ and $ \derLmu `G |- M : `k_i'\arrow`n | `a{:}`d_i\prod `k_i,`b{:}`k_i',`D' $.
We can then construct:
 \[
 \Inf	[\App]
	{\Inf	[`m]
		{\InfBox{ \derLmu `G |- M : `k_i'\arrow`n | `a{:}`d_i\prod `k_i,`b{:}`k_i',`D' }
		}{\derLmu `G |- `m`a.[`b]M : `d_i\prod `k_i\arrow`n | `D }
	 \InfBox{ \derLmu `G |- N : `d_i | `D }
	}{\derLmu `G |- (`m`a.[`b]M)N : `k_i\arrow`n | `D }
 \]
As above, we conclude that $ \derLmu `G |- (`m`a.[`b]M)N : `d | `D $ by rule $(\inters)$.
 \QED

 \end{description}
 \end{enumerate}
 \end{proof}
 \end{lemma}

We will now show that all strongly normalisable terms are typeable in our system.
The proof of the crucial lemma for this result as presented below (Lemma~\ref {l.o. lemma}) goes by induction on the left-most outer-most reduction path.

 \begin {definition}
An occurrence of a redex $\Redex = ( `l x.P)Q$ or $(`m`a.[`b]P)Q$ in a term $M$ is called the \emph{left-most outer-most redex of $M$} ($\Lor(M)$), if and only if:

 \begin {enumerate}
 \item there is no redex $\Redex'$ in $M$ such that $\Redex' = \Cont [ \Redex ] $ (\emph{outer-most});
 \item there is no redex $\Redex'$ in $M$ such that $M = \Cont_0 [{ \Cont_1 [ \Redex' ] \, \Cont_2 [ \Redex ] }] $ (\emph{left-most}).
 \end {enumerate}
$M \lored N$ is used to indicate that $M$ reduces to $N$ by contracting $\Lor(M)$.

 \end {definition}


The following lemma formulates a subject expansion result for our system with respect to left-most outer-most reduction.
A proof for this property in the context of strict intersection type assignment for the $`l$-calculus appeared in \cite{Bakel-NDJFL'04,Bakel-ACM'11}.

 \begin {lemma} \label {l.o. lemma}
Let $M \lored N$, $\Lor(M) = RQ$, $\derLmu `G_1 |- N : `d_1 | `D_1 $ with $`d_1$ not an intersection, and $\derLmu `G_2 |- Q : `d_2 | `D_2 $, then there exist $`G_3$, $`D_3$ and $`d_3$ such that $`G_3\leq `G_1$, $`D_3\leq `D_1$, $ `d_1 \leq `d_3$, and 
$\derLmu `G_3 |- M : `d_3 | `D_3 $.
 
\Comment{
 \[ \begin{array}{lrcll}
\Lang_D : & `d,`c,`j &::=& `n \mid `w\arrow`n \mid `k\arrow`n \mid `d \inter `d \\
\Lang_C : & `k,`p,`q &::= & `d\prod `w \mid`d\prod `k \mid `k \inter `k
 \end{array} \]
}

 \begin{proof}
By induction on the structure of terms.

 \begin{description}

 \item[$M = VP_1\dots P_n :$]
Then either:
 \begin {enumerate}

 \item 
$V$ is a redex $( `l y.P)Q$, so $\Lor(M) = V$; let $ V' \same P [ Q/y ]$; or 

 \item 
$V$ is a redex $(`m`a.[`b]P)Q$, so $\Lor(M) = V$; let $ V' \same `m`a.[`b] P \StrSub{`a }{Q} $; or

 \item 
$V\same z$ and there is an $i \ele \n$ such that $\Lor(M) = \Lor(P_j) $, $N \same zP_1\dots P'_i\dots P_n $, and $P_i \lored P_i'$; let $ V' = z$.

 \end {enumerate}
By assumption $`d_1 = `k_1\arrow`n$.
Then there are $ `d_j~(j \ele \n)$, such that $ \derLmu `G_1 |- V' :  `d'_{1}\prod \dots \prod `d'_{n}\prod `k_1\arrow`n | `D_1 $ and $ \derLmu `G_1 |- P_i : `d'_i | `D_1 $, for all $i \ele \n $.

%

We distinguish:
 \begin {enumerate}

 \item $V' \same P [ Q/y ] $, where the substitution is capture avoiding, so all free variables in $Q$ are free in $P [ Q/y ]$ when $y \ele \fv(P)$, and we can assume that $`G_2$ and $`D_2$ do not have types for bound variables and names in $P$.
Let $`G_3 = `G_1 \inter `G_2$ and $`D_3 = `D_1 \inter `D_2$, then by Corollay~\ref{lem:contextInters} and Lemma~\ref {lem:redexExpMu},
 $
\derLmu `G_3 |- (`ly.P)Q : `d'_{1}\prod \dots \prod `d'_{n}\prod `k_1\arrow`n | `D_3
 $.

 \item $V' \same `m`a.[`b] P \StrSub{`a }{Q} $; we can assume that $`G_2$ and $`D_2$ do not have types for bound variables and names in $`m`a.[`b] P$.
Let $`G_3 = `G_1 \inter `G_2$ and $`D_3 = `D_1 \inter `D_2$, then by Corollay~\ref{lem:contextInters} and Lemma \ref {lem:redexExpMu},
 $
\derLmu `G_3 |- (`m`a.[`b] P ) Q : `d'_{1}\prod \dots \prod `d'_{n}\prod `k_1\arrow`n | `D_3 
 $.

 \item $V' \same z$.
Then, by induction, there are $`G'$, $`D'$, $`d''_j$ such that $ `d''_j \leq `d'_j $, and 
$\derLmu `G' |- P_j : `j'_j | `D' $.
Take $`G_3 = `G_1 \inter `G', z{:}`d'_1\prod \dots \prod `d''_j\prod \dots \prod `d'_n \prod `k'_1 \arrow `n $, and $`D_3 = `D_1 \inter `D'$, then
 \[
\derLmu `G_3 |- z : `d'_1\prod \dots \prod `d''_j\prod \dots \prod `d'_n \prod `k'_1 \arrow `n | `D_3
 . \]

 \end {enumerate}

In all cases, $`G_3 \leq `G_1$, $`D_3 \leq `D_1$, and $ \derLmu `G_3 |- VP_1\dots P_n : `d | `D_3 $.

 \item[$M = `ly.M' :$]
If $M \lored N$, then $N = `ly.N'$ and $M' \lored N'$.
Then there exists $`d$ and $`k$ such that $`d_1 = `d\prod `k \arrow `n$, and $\derLmu `G_1,y{:}`d |- N' : `k\arrow `n | `D_1 $.
By induction, there exists $`G' \leq `G_1$, $`D' \leq `D_1$, $`d' \leq `d$, and $`k'\leq `k$ such that $\derLmu `G',y{:}`d' |- M' : `k'\arrow `n | `D' $.
Then, by rule $(\Abs)$, $\derLmu `G' |- `ly.M' : `d'\prod `k'\arrow `n | `D' $.
Notice that $`d\prod `k \arrow `n \leq `d'\prod `k'\arrow `n$; take $`G_3 = `G'$, $`D_3 = `D'$, and $`d_3 = `d'\prod `k'\arrow `n$.

 \item[{$M = `m`a.[`b]M' :$}]
If $M \lored N$, then $N = `m`a.[`b]N'$ and $M' \lored N'$.
Then there exists $`k_1$ and $`k_2$ such that $`d_1 = `k_1 \arrow `n$, $`D_1 = `a{:}`k_2,`D'_1$ and $\derLmu `G_1 |- N' : `k_2\arrow `n | `b{:}`k_1,`D'_1 $.
By induction, there exists $`G' \leq `G_1$, $`D' \leq `D_1$, $`k'\leq `k_1$ and $`k'' \leq `k_2$ such that $\derLmu `G' |- M' : `k''\arrow `n | `a{:}`k',`D' $.
Then, by rule $(`m)$, $\derLmu `G' |- `m`a.[`b]M' : `k'\arrow `n | `b{:}`k'',`D' $.
Notice that $`k_1 \arrow `n \leq `k'\arrow `n$, and $`b{:}`k'',`D' \leq `b{:}`k_2,`D'_1 $; take $`G_3 = `G'$, $`D_3 = `b{:}`k'',`D'$, and $`d_3 = `k'\arrow `n$.

 \item[{$M = `m`a.[`a]M' :$}]
If $M \lored N$, then $N = `m`a.[`a]N'$ and $M' \lored N'$.
Then there exists $`k$ such that $`d_1 = `k \arrow `n$, $`D_1 = `a{:}`k,`D'_1$ and $\derLmu `G_1 |- N' : `k\arrow `n | `a{:}`k,`D'_1 $.
By induction, there exists $`G' \leq `G_1$, $`D' \leq `D_1$, and $`k_1\leq `k$, $`k_2\leq `k$ such that $\derLmu `G' |- M' : `k_2\arrow `n | `a{:}`k_1,`D' $.
Take $`k' = `k_1 \inter `k_2$, then by weakening and rule $(\leq)$, also $\derLmu `G' |- M' : `k'\arrow `n | `a{:}`k',`D' $.
Then, by rule $(`m)$, $\derLmu `G' |- `m`a.[`a]M' : `k'\arrow `n | `D' $.
Notice that $`k \arrow `n \leq `k'\arrow `n$; take $`G_3 = `G'$, $`D_3 = `D'$, and $`d_3 = `k'\arrow `n$.
\QED

 \end{description}
 \end{proof}
 \end {lemma}

 \Comment{
With this result, the following property now follows easily.

 \begin{lemma}[Subject Expansion] \label{lem:expansion}
If $M \reduces N$ by contracting either a $`b$-redex $(`lx.P)Q$ or a $`m$-redex
$(`m`a.[`b]P)Q$ and $ \derLmu `G |- Q : `d' | `D $ for some $`d'$, then $ \derLmu `G |- N : `d | `D $ implies $ \derLmu `G |- M : `d | `D $.

 \begin{proof}
By induction over the structure of derivations, using Lemma~\ref{lem:redexExpBeta}.\QED
 \end{proof}
 \end{lemma}
}

We can now show that all strongly normalisable terms are typeable in our system.

 \begin{theorem}[Typeability of $ \SN$-Terms]
For all $M \ele \SN$ there exist $`G$ and $`D$ and a type $`d$ such that $ \derLmu `G |- M : `d | `D $.
 
 \begin{proof}
By induction on the maximum of the lengths of reduction sequences for a strongly normalisable term to its normal form (denoted by $\#(M)$).

 \begin {enumerate}
 \item
If $\#(M) = 0$, then $M$ is in normal form, and by Lemma~\ref{from normal form}, there exist $`G$ and $ `d$ such that $\derLmu `G |- M : `d | `D $.

 \item
If $\#(M)\geq 1$, so $M$ contains a redex, then let $M\lored N$ by contracting $PQ$.
Then $\#(N) < \#(M)$, and $\#(Q) < \#(M)$ (since $Q$ is a proper subterm of a redex in $M$), so by induction $\derLmu `G |- N : `d_1  | `D $ and $\derLmu `G' |- Q : `d_2  | `D $, for some $`G$, $`G'$, $`d_1$, and $ `d_2$.
Then, by Lemma~\ref {l.o. lemma}, there exist $`G_1$, $`D_1$ and $ `d'$ such that $\derLmu `G_1 |- M : `d'  | `D_1 $.\QED

 \end{enumerate}
 \end{proof}
 \end{theorem}

In the following section we will prove strong normalisation for terms typeable in the propositional fragment of Parigot's logical system \cite{Parigot'92} via an interpretation in our system.

 \section{Interpretation of Parigot's Logical System}\label{sec:Parigot}

We use a version of Parigot's logical system (as presented in \cite{Parigot'92} which is equivalent to the original one if only terms (so not also proper commands, i.e.~elements of $\Commands$) are typed. This implies that the rule for $\bot$ does not need to be taken into account.%
\footnote{The system we consider here does not include rules $(\quaI)$ and $(\quaE)$, since they have no effect on the subject in Parigot's first-order type assignment system.} 
We call this propositional fragment of Parigot's original system the \emph{simply-typed $`l`m$-calculus}.

 \begin{definition}[Simply Typed $`l`m$-calculus] 
 \begin{enumerate}

 \item
The set $\Formulas$ of {\em Logical Formulas} is defined by 
 \[A,B ::= \varphi \mid A \arrow B\]
where $\varphi$ ranges over an infinite set of \emph{Proposition (Type) Variables}.


 \item
The inference rules of this system are:
 \[ \begin{array}{rl@{\quad}rl@{\quad}rl}
(\Ax) : &
 \Inf	
	{ \derLmu `G , x{:}A |- x : A | `D }
&
(`m_1): &
 \Inf	
	{ \derLmu `G |- M : A | `a{:}A,`D
	}{ \derLmu `G |- `m`a.[`a]M : A | `D }
&
(`m_2): &
 \Inf	
	{\derLmu `G |- M : B | `a{:}A, `b{:}B,`D
	}{\derLmu `G |- `m`a[`b]M : A | `b{:}B,`D }
 \end{array} \]
 \[ \begin{array}{rl@{\quad}rl}
(\arrI): &
 \Inf	
	{ \derLmu `G,x{:}A |- M : B | `D
	}{ \derLmu `G |- `l x.M : A\arrow B | `D }
&
(\arrE): &
 \Inf	
	{\derLmu `G |- M : A\arrow B | `D
	 \quad
	 \derLmu `G |- N : A | `D
	}{\derLmu `G |- MN : B | `D }
 \end{array} \]

We write $ \derLmuP `G |- M : A | `D $ to denote that this judgement is derivable in this system.

 \end{enumerate}
 \end{definition}

We can interpret formulas into types of our system as follows.

 \begin{definition}\label{def:translation}
The translation functions $(\cdot)^D:\Formulas \arrow \Lang_D$ and $(\cdot)^C:\Formulas \arrow \Lang_C$~ are defined by (remember that $`n$ is the (only) base type): 
 \[ \begin{array}{rcl}
 \varphi^C &=& `n\prod`w
	\\
(A\arrow B)^C &=& (A^C\arrow`n)\prod B^C
	\\
A^D &=& A^C\arrow `n
 \end{array} \]
 \end{definition}
For example, 
$(`v_1 \arrow `v_2 \arrow `v_3)^C = 
(`n\prod`w \arrow `n) \prod (`n\prod`w \arrow `n) \prod (`n\prod`w \arrow `n) $.

It is straightforward to show that the above translations are well defined.
We extend them to bases and name contexts as follows: $`G^D = \Set{x{:}A^D \mid x{:}A \ele `G}$ and $`D^C = \Set{`a{:}A^C \mid `a{:}A \ele `D}$. 

 \Comment{
 \begin{lemma}\label{lem:translation}
The translation is well defined. More precisely, given a formula $A$,
we have that $A^D\in\Lang_D$ and $A^C\in\Lang_C$.

 \begin{proof} By straightforward induction over $A$.
 \QED
 \end{proof}
 \end{lemma}
}

 \begin{theorem}[Derivability preservation]\label{thr:translation}
If $\derLmuP `G |- M : A | `D $, then $ \derLmu `G^{D} |- M : A^{D} | `D^{C} $.
 \begin{proof} 
By induction on the structure of derivations.
Each rule of the simply-typed $`l`m$-calculus has a corresponding one in our intersection type system (allowing for the fact that rule $(\arrI)$ gets mapped unto $(\Abs)$ and $(\arrE)$ gets mapped unto $(\App)$); hence it suffices to show that
rules are preserved when translating formulas into types. We show just the cases for the $`m$-abstraction.
 \Comment{
 \noindent Case $(\mbox{\it Ax})$: the axiom becomes:
 \[
 \Inf (\mbox{\it Var}]{\derLmu `G^D,x{:}A^D |- x : A^D | `D^C }

 \noindent Case $(\arrow \mbox{\it I})$: we translate $A^D = A^C\arrow`n$, $B^D = B^C\arrow`n$ and
 $(A\arrow B)^D = (A\arrow B)^C\arrow`n = ((A^C\arrow`n)\prod B^C)\arrow`n$. Therefore we have:
 \[ \begin{array}{ccc}
 \Inf	{\derLmu `G^D, x{:}A^D|- M : B^D | `D^C
	}{ \derLmu `G^D |- `l x.M : (A\arrow B)^D | `D^C }
& = &
 \Inf	[\arrI]
	{\derLmu `G^D, x{:} A^C\arrow`n |- M : B^C\arrow`n | `D
	}{\derLmu `G^D |- `l x.M : ((A^C\arrow`n)\prod B^C)\arrow`n | `D }
 \end{array}
 \]
 \noindent Case $(\arrow \mbox{\it E})$:
 \[ \begin{array}{cc}
 \Inf	{\derLmu `G^D |- M : (A\arrow B)^D | `D^C 
	 \quad
	 \derLmu `G |- N : A^D | `D^C
	}{\derLmu `G |- MN : B^D | `D }
 \\ [6mm]
= &
 \Inf	[\arrE]
	{\derLmu `G^D |- M : ((A^C\arrow`n)\prod B^C)\arrow`n | `D^C 
	 \quad
	 \derLmu `G |- N : A^C\arrow`n | `D^C
	}{\derLmu `G^D |- MN : B^C\arrow`n | `D^C }
 \end{array} \]
}

 \[ \begin{array}[t]{ccc}
 \Inf	[`m_1]
	{\InfBox {\derLmu `G^D |- M : A^D | `a{:}A^C,`D^C }
	}{\derLmu `G^D |- `m`a.[`a]M : A^D | `D^C }
& \textrm{becomes} &
 \Inf	[`m]
	{\InfBox {\derLmu `G^D |- M : A^C\arrow`n | `a{:}A^C,`D^C }
	}{\derLmu `G^D |- `m`a.[`a]M : A^C\arrow`n | `D^C }
 \\ [8mm]
 \Inf	[`m_2]
	{\InfBox {\derLmu `G^D |- M : B^D | `a{:}A^C,`b{:}B^C,`D^C }
	}{\derLmu `G^D |- `m`a.[`b]M : A^D | `b{:}B^C,`D^C }
& \textrm{becomes} &
 \Inf	[`m]
	{\InfBox {\derLmu `G^D |- M : B^C\arrow`n | `a{:}A^C,`b{:}B^C,`D^C }
	}{\derLmu `G^D |- `m`a.[`b]M : A^C\arrow`n | `b{:}B^C,`D^C }
 \end{array} \] 
notice that the applications of rule $(`m)$ are valid instances of that rule.\QED

 \end{proof}
 \end{theorem}

Strong normalisation of typeable terms in Parigot's simply typed $`l`m$-calculus 
now follows as a consequence of our characterisation result.

 \begin{theorem}[Strong Normalisability of Parigot's Simply Typed $`l`m$-calculus] \label{cor:translation} 
If $\derLmuP `G |- M : A | `D $, then $M\ele \SN$.

 \begin{proof} By Theorem \ref{thr:translation}, if $\derLmuP `G |- M : A | `D $ then 
$\derLmu `G^D |- M : A^D | `D^C $ is derivable in the intersection type system. Hence $M\in\SN$ by Theorem
 \ref{thr:typableAreSN}.
 \QED
 \end{proof}
 \end{theorem}

 \section*{Conclusion}
We have defined an intersection type system which characterises strongly normalising $`l`m$-terms, extending the strong normalisation result for the $`l$-calculus to the pure $`l`m$-calculus. 

We have also provided a translation of propositional types of Parigot's system into types of the system proposed in this paper (a restriction of the one presented in \cite{Bakel-Barbanera-Liguoro-TLCA'11}) and proved that derivability is preserved. 
We are confident that such a result can be extended to the full first-order type assignment system, to obtain an alternative proof of Parigot's strong normalisation theorem.

As we have observed in \cite{Bakel-Barbanera-Liguoro-TLCA'11}, our intersection-type assignment system can be adapted to de Groote's variant of the $`l`m$-calculus (see e.g.~\cite{deGroote'94}) 
(called $`L`m$ by Saurin \cite{Saurin'08}), that satisfies stronger properties than Parigot's original calculus, such as B\"ohm's theorem. 
We leave the question whether the present characterisation result extends to those cases to future work.

\paragraph*{Acknowledgements}
The authors wish to thank Mariangiola Dezani-Ciancaglini for her unceasing support.

 \bibliography{references}

 \end{document}